\documentclass[10pt,twocolumn]{article}

\usepackage[]{graphicx}
\usepackage{graphicx,amsmath,color}
\usepackage{txfonts}
\usepackage{multirow}
\usepackage{subfigure}
\usepackage{natbib}
\usepackage{url}
\bibpunct{(}{)}{;}{a}{}{,} % to follow the A&A style
\begin{document}

\title{Apodized phase mask coronagraphs for arbitrary apertures}

\author{A. Carlotti$^{1}$}

\date{ \small{$^{1}$ HCIL, Mechanical \& Aerospace Engineering dpt., Princeton University, Princeton, NJ, 08544} \\  \vspace{1mm} email: \url{acarlott@princeton.edu} \\ \vspace{2mm} \normalsize{December 26, 2012}}

% \abstract{}{}{}{}{} 
% 5 {} token are mandatory

  % \keywords{Instrumentation: high angular resolution -- Techniques: high angular resolution -- Methods: analytical -- Methods: numerical}

   \maketitle

%
%________________________________________________________________

  \abstract{Phase masks coronagraphs can be seen as linear systems that spatially redistribute, in the pupil plane, the energy collected by the telescope. Most of the on-axis light must ideally be rejected outside the aperture so as to be blocked with a Lyot stop, while almost all of the off-axis light must go through it. The unobstructed circular apertures of off-axis telescopes make this possible but all of the major telescopes are however on-axis and the performance of these coronagraphs is dramatically reduced by their central obstructions.\\
  Their performance can be restored by using an additional optimally designed apodizer that changes the amplitude in the first pupil plane so that the on-axis light is rejected outside the obstructed aperture of the on-axis telescope. \\
  An apodizer is assumed to be located in a pupil plane, a phase mask in a subsequent image plane, and a Lyot stop in a reimaged pupil plane. The numerical optimization model is built by maximizing the apodizer's transmission while setting constraints on the extremum values of the electric field that the Lyot stop does not block. The coronagraphic image is compared to what a non-apodized phase mask coronagraph provides and an analysis is made of the trade-offs that exist between the transmission of the apodizer and the properties of the Lyot stop. \\
  The existence of a solution and the transmission of the mask depend on the geometries of the aperture and of the Lyot stop, and on the constraints that are set on the on-axis attenuation. The system's throughput is a concave function of the Lyot stop transmission. In the case of a VLT-like aperture, optimal apodizers with a transmission of 16\% to 92\% associated with a four-quadrant phase mask provide contrast as low as a few $10^{-10}$ at $1 \, \lambda/D$ from the star. The system's maximum throughput is about 64\%, for an apodizer with an 88\% transmission and a Lyot stop with a 69\% transmission. It is showed that optimizing apodizers for a vortex phase mask requires computation times much longer than in the previous case, and no result is presented for this mask. \\
  It is demonstrated that apodizers can be successfully optimized to allow phase mask coronagraphs to be used with the full aperture of on-axis telescopes while delivering contrast as low as or even lower than what they could provide by themselves with off-axis telescopes.}

\section{Introduction}

Over the past 20 years, indirect detection methods have provided us with a now exhaustive catalogue of confirmed companions (see for example \url{http://exoplanet.eu/catalog/}) and an even greater list of potential companions (for the Kepler candidates, see \url{http://archive.stsci.edu/kepler/planet_candidates.html}). One of the most striking conclusion so far is that the fraction of rocky planets (including larger and/or heavier Earths) is larger than previously anticipated\citep{Borucki2011,Batalha2012}. It is however only valid for planets with short periods, thus at a small distance from their star (about 0.25 AU in the case of a solar-type star).

Several indirect detection methods exist: radial velocimetry, transits, astrometry, micro-lensing, pulsar timing. Transits have been used to probe exoplanets atmospheres \citep{Burrows2006}, and the presence of basic molecules such as water, methane, or carbon dioxide has been detected in the atmosphere of 5 planets. 

While direct imaging methods cannot currently be used to characterize the planets detected by transits, they can be used to look for larger planets, further away from their star. Some instruments are already being used \citep{Hinkley2011}.

The spatial and spectral resolutions of near-future direct imaging near-infrared instruments will make it possible to study planets around nearby stars in much finer details \citep{Macintosh2008,Beuzit2010,Beichman2010,Lagage2010,Sivaramakrishnan2010,Enya2011}.

Many instruments have been proposed to detect and analyze planet light by direct imaging. In this paper, phase mask coronagraphs and apodized coronagraphs are primarily considered. The stellar Lyot coronagraph \citep{Lyot1932,Kenknight1977} was first modified by Roddier \& Roddier \citep{Roddier1997}. The focal amplitude mask was replaced by a focal phase mask, a disk centered on the star that induces a $\pi$ phase shift. Destructive interferences in the following pupil plane lead to a partial cancellation of the star light. It was shown later that this null can be made total if a spheroidal prolate apodizer is used in addition \citep{Aime2002}.

Rouan proposed another phase mask, the four quadrant phase mask (4QPM) to cancel the on-axis starlight entirely, without using any apodization \citep{Rouan2000}. This mask induces a $\pi$ phase shift of the electric field in two opposite quadrants of the image plane, attenuating slightly the light coming from the planets if it happens to fall on the edges of the quadrants. Eight octants phase mask (8OPM) have also been studied. While their geometry reduces the discovery area in the image plane, they are also less sensitive to low-order aberrations \citep{Murakami2008,Carlotti2009} than the 4QPM.

These phase mask coronagraphs may suffer from chromaticity: $\pi$ phase shifts are usually induced by laying on the surface of a plane-parallel window a thin transparent film that is then selectively etched \citep{Riaud2003}. The phase shift that is induced depends on the thickness of the film, its refraction index and on the wavelength. A first solution to this problem is to use N masks designed as many different wavelengths in as many successive image planes \citep{Galicher2011}. Other possibilities to induce a partial achromatic phase shift consist in using semi-achromatic polarizing elements (half-wave plates) \citep{Mawet2003, Boccaletti2008}, or sub-wavelength gratings \citep{Mawet2005}. A Mach-Zehnder interferometer can also be used to induce an achromatic $\pi$ phase shift \citep{Aime2007}, although many more optical elements are required than in the other designs.

Finally, \cite{Mawet2005} presented the vector vortex coronagraph (VVC). The vortex phase mask (VPM) induces an azimuthal phase shift, from 0 to $2 l \theta$, where $\theta$ is the azimuth, and $l$ the topological charge, an even integer strictly greater than 0. The VVC also cancels the on-axis star light entirely, but contrary to the 4QPM, the geometry of the VPM does not reduce the discovery area. Several manufacturing techniques are considered for the VVC (liquid crystal polymers, sub-wavelength gratings, and photonics crystals), and a partial achromatization is possible (for more details see \cite{Mawet2011}).

The Roddier \& Roddier mask can be modified to improve its overall performance and its achromaticity by adding a phase-shifting annulus to the central mask to form a dual-zone mask \citep{Soummer2003}. Like the Roddier \& Roddier mask, the dual-zone mask can be combined with an apodization, which can induce a perfect null in a broadband as long as its transmission varies chromatically in a specific manner \citep{N'diaye2012}.

All phase masks have been extensively studied through numerical simulations, and many of them through laboratory experiments \citep{Mawet2009,Baudoz2010}. On-sky observations have been conducted with the 4QPM coronagraph \citep{Gratadour2005} and the VVC \citep{Mawet2010}, however all three masks have been designed for circular clear apertures, not centrally obstructed apertures. As a consequence their performance has been greatly limited. The VVC has for example been tested on a 1.5m subaperture of the 5m Palomar telescope, reducing its effective resolution to a third of its value, and its transmission to a ninth. It was nevertheless able to observe successfully 3 of the 4 planets of the HR8799 planetary system \citep{Serabyn2010} (more information on the discovery of these planets can be found in \cite{Marois2008,Marois2010}). In the specific case of this observation, using a subaperture also had the advantage of mitigating the impact of wavefront errors.

Phase masks coronagraphs will be used in three major high-contrast instruments. On the ground, the spectro-polarimetric high-contrast exoplanet research (SPHERE) instrument will use three 4QPM designed for the J, H and K bands \citep{Beuzit2010}. In space, the mid-infrared instrument of the James Webb space telescope (JWST-MIRI) will also use a series of 4QPM \citep{Cavarroc2008}. SPHERE should see its first light in 2013, while the JWST should be launched in 2018. A partially transmissive Roddier \& Roddier mask is currently studied for the SCExAO instrument at the Subaru telescope \citep{Martinache2011}, where it would be used in the focal plane of a phase-induced amplitude apodization (PIAA) coronagraph \citep{Guyon2003,Guyon2010}. Still at the Subaru telescope, 4QPM, 8OPM \citep{Murakami2010} and VPM are also likely to be tested.

A possible solution to the problem of on-axis observations, apart from the use of subapertures, involves two-stages optical layouts. For instance, two VPM \citep{Serabyn2011} or two 4QPM \citep{Galicher2011} can be set in series (in two successive image planes). This reduces the impact of the central obstruction. It also involves twice as many optics, each of them bringing additional aberrations to the wavefront, and making the alignment more difficult than is a single stage layout.

This paper presents a different solution to on-axis observations with phase mask coronagraphs. Apodizers can be optimally designed in two dimensions to cancel the on-axis light of a star when used with a 4QPM or VPM. The apodizer is set in a pupil plane that comes before the image plane in which the phase mask is located. The optimization of the apodizer is in many ways similar to the optimization of 2 dimensional binary apodizers, i.e. 2D shaped pupils. 

While binary pupil masks have long been used by astronomers \citep{Schultz1983}, their optimization for high-contrast has however first been proposed about ten years ago \citep{Spergel2001}. Several different types of masks have been designed since \citep{Vanderbei2003, Kasdin2007}, and they share properties with external occulters. While initially optimized in one dimension, it was shown more recently that they can be optimized in two dimensions \citep{Carlotti2011}. This makes it possible to tailor the pupil plane transmission pixel by pixel. As a consequence apodizers can be computed for any aperture, including centrally obstructed and segmented apertures.

The optimization problem consists in maximizing the transmission of the apodizer under constraints set on the extremum values of the amplitude of the electric field in chosen regions of the image plane. Because the problem is convex and the electric field is computed through a linear operation (a Fourier transform), a unique solution exists and can be found efficiently. To limit the computation time required by these numerical optimizations, the number of resolution elements along both axes of the pupil plane is usually kept under a thousand, and under a few tens in the image plane.

Computing a shaped pupil that restores the performance of a phase mask coronagraph is also possible, though at first glance the complexity of the computations appears to be much greater since one or two more Fourier transforms must be computed. Constraints are set on the extremum values of the amplitude of the electric field, either in the Lyot plane (within the limits of a Lyot stop) or in the final image plane.

The problem can however be simplified by noting that the operator that transforms the electric field in the first pupil plane into the electric field in the second pupil plane is linear. It is thus not mandatory to explicitly compute the two successive Fourier transforms. This can drastically reduce the complexity of the computation. More importantly, it prevents computation errors to spread from one plane to another and eliminates the constraint on the sampling of the electric field in the first image plane. 

Even if the electric field in the intermediate image plane is not explicitly computed, the finite angular extension of the image plane must be specified. Phase mask coronagraphs require the phase shift to be applied on the largest possible area of the image plane, usually within a circle with a minimum radius of $32$ or $64 \lambda/D$ \citep{Riaud2001}.

%While wavefront aberrations are not considered in this paper, any high-contrast imaging instrument must be used with an extreme adaptive optics system. In practice, deformable mirrors are used to control the amplitude and the phase of the wavefront within the limits of a control area. Its width depends on the number of actuators, and currently does not exceed 64 $\lambda/D$.

Sec. \ref{Maths} derives the equations that are later used to compute the electric field in the Lyot plane as a function of the apodizing pattern used in the entrance pupil plane, and of the phase-shift induced by the mask in the first pupil plane. The case of the 4QPM and of the VPM are considered separately. Section \ref{DiscretizationOptimization} gives the details of the discretization of the calculations and the construction of the optimization problem. Sec. \ref{Simulations} analyzes the impacts of the central obstruction and of the finite size of the focal mask on the high-contrast performance of a 4QPM and a VPM. The results of the optimizations are presented in the same section, before being discussed in Sec. \ref{Conclusion}.

\section{Fourier optics formalism}
\label{Maths}

\begin{figure}[t]
\centerline{\includegraphics[width=1.0\columnwidth]{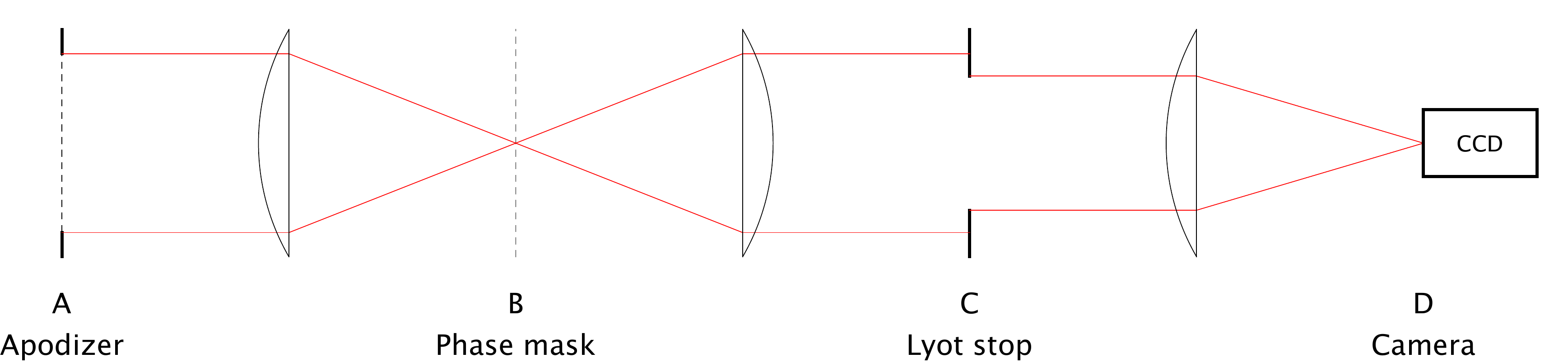}}
\caption{Schematic drawing of the optical layout. The successive pupil and image planes A, B, C and D are the locations of the apodizer, the phase mask, the Lyot stop and the camera. The plano-convex lenses were chosen to denote the use of a focusing element. They could for example be replaced with achromatic doublets, spherical or off-axis parabolic mirrors.}
\label{Layout}
\end{figure}

The optical layout is represented in Fig. \ref{Layout}. Four planes are indicated, noted with letters A,B,C and D. They correspond to the successive pupil and image planes, where the apodizer, the phase mask, the Lyot stop and the camera are respectively located. The telescope's pupil is assumed to have a diameter D. The observation wavelength is noted $\lambda$, and the letter F denotes the focal length of the lenses or mirrors used to reimage the various pupil and image planes.

The apodizer and the telescope aperture are assumed to be symmetric with respect to the x and y axes, and since their transmission is real, the Fourier transform $E(u,v)$ of the apodizer A(x,y) is a real, even function (with the exception of a scalar uniform phase factor). Its expression is given as follows, in angular units:
\begin{equation}
\begin{split}
E(u,v) & = \frac{e^{2 i \pi F / \lambda}}{i \lambda F} \, \iint_{-D/2}^{D/2} A(x,y)\, e^{-2 i \pi \, (u x + v y)} \, dx \, dy \\
          & = 4 \frac{e^{2 i \pi F / \lambda}}{i \lambda F} \, \iint_{0}^{D/2} A(x,y)\, \cos \big( 2 \pi u x \big) \, \cos \big( 2 \pi v y \big) \, dx \, dy.
\end{split}
\end{equation}
\subsection{Case of the four-quadrant phase mask}
The 4QPM can be described through the odd function $M_{FQ}(u,v)=sign(u) \times sign(v)$, and the electric field in the reimaged pupil plane is the inverse Fourier transform of $E(u,v) \times M_{FQ}(u,v)$:

\begin{equation}
\begin{split}
P(\tilde{x},\tilde{y})=-i \lambda F \, e^{2 i \pi F / \lambda} &\iint_{-\infty}^{\infty} E(u,v) \times M_{FQ}(u,v) \, e^{2 i \pi (u \tilde{x} + v \tilde{y})} du\, dv \\
	 =-i \lambda F \, e^{2 i \pi F / \lambda} \, \Big( &\int_{0}^{\infty} \int_{0}^{\infty} E(u,v) \, e^{2 i \pi (u \tilde{x} + v \tilde{y})} du\, dv \\
	 &-\int_{-\infty}^{0} \int_{0}^{\infty} E(u,v) \, e^{2 i \pi (u \tilde{x} + v \tilde{y})} du\, dv \\
	 &-\int_{0}^{\infty} \int_{-\infty}^{0} E(u,v) \, e^{2 i \pi (u \tilde{x} + v \tilde{y})} du\, dv \\
	 &+\int_{-\infty}^{0} \int_{-\infty}^{0} E(u,v) \, e^{2 i \pi (u \tilde{x} + v \tilde{y})} du\, dv \Big) \\
\end{split}
\end{equation}
%
%\begin{equation}
%\begin{split}
%P(x,y)=&\int_{0}^{\infty} \int_{0}^{\infty} E(u,v) \, e^{2 i \pi (u x + v y)} du\, dv \\
%	 -&\int_{0}^{\infty} \int_{0}^{\infty} E(-u,v) \, e^{-2 i \pi (u x - v y)} du\, dv \\
%	 -&\int_{0}^{\infty} \int_{0}^{\infty} E(u,-v) \, e^{2 i \pi (u x - v y)} du\, dv \\
%	 +&\int_{0}^{\infty} \int_{0}^{\infty} E(-u,-v) \, e^{-2 i \pi (u x + v y)} du\, dv 
%\end{split}
%\end{equation}
%
%\begin{equation}
%\begin{split}
%P(x,y) =\iint_{0}^{\infty} & E(u,v) \, \Big(e^{2 i \pi (u x + v y)}+e^{-2 i \pi (u x + v y)}\\
%	 &-\big( e^{2 i \pi (u x - v y)} +e^{-2 i \pi (u x - v y)} \big) \Big)\, du\, dv\\
%= 2 \iint_{0}^{\infty} & E(u,v) \, \Big(\cos \big(2 \pi (u x + v y)\big)\\
%	 &-\cos \big(2 \pi (u x - v y)\big) \Big)\, du\, dv
%\end{split}
%\end{equation}

For the moment, the dimensions of the focal phase mask are assumed to be infinite. After a few algebraic transformations, this expression can be reduced to:

\begin{equation}
P(\tilde{x}, \tilde{y}) = 4 i \lambda F \, e^{2 i \pi F / \lambda} \iint_{0}^{\infty} E(u,v) \, \sin \big( 2 \pi u \tilde{x} \big) \sin \big( 2 \pi v \tilde{y} \big)\, du\, dv
\end{equation}

Plugging in the expression for E(u,v) as a function of A(x,y) leads to:

\begin{equation}
\begin{split}
P(\tilde{x},\tilde{y}) = 16 \, e^{4 i \pi F/\lambda} \iint_{0}^{\infty} & \iint_{0}^{D/2} A(x,y) \, \cos \big( 2 \pi u x \big) \, \cos \big( 2 \pi v y \big) \\ 
& \times \, \sin \big( 2 \pi u \tilde{x} \big) \sin \big( 2 \pi v \tilde{y} \big)\, dx \, dy \, du\, dv 
\end{split}
\end{equation}

The integrations over $u$ and $v$ can be considered separately:

\begin{equation}
\begin{split}
P(\tilde{x},\tilde{y}) = 16 \, e^{4 i \pi F/\lambda} & \iint_{0}^{D/2} A(x,y) \, \Big[\int_{0}^{\infty} \cos \big( 2 \pi u x \big) \sin \big( 2 \pi u \tilde{x} \big) \, du \\
& \times \int_{0}^{\infty} \cos \big( 2 \pi v y \big) \sin \big( 2 \pi v \tilde{y} \big) \, dv \Big ] \, dx \, dy
\end{split}
\end{equation}

or equivalently:

\begin{equation}
P(\tilde{x},\tilde{y}) = 16 \, e^{4 i \pi F/\lambda} \iint_{0}^{D/2} A(x,y) \, C_{\infty}(x,\tilde{x}) \, C_{\infty}(y,\tilde{y}) \, dx \, dy  
\label{P2P}
\end{equation}

where the function $C_{\infty}(\alpha,\beta)$ that has been introduced is defined as:

\begin{equation}
\begin{split}
C_{\infty}(\alpha,\beta) &= \int_{0}^{\infty} \cos \big( 2 \pi u \alpha \big) \sin \big( 2 \pi u \beta \big) \, du
\end{split}
\end{equation}

While it is not possible to find an analytical expression for $C_{\infty}$ if the integration domain is semi-infinite, it becomes possible for a finite domain. In fact phase masks can only cover a finite area of the image plane, and this restriction - instead of being an artificial assumption - is mandatory. It also means that any light outside this finite domain is blocked. This loss of throughput is however negligible (for example, more than 99\% of the Airy pattern's energy is concentrated inside a circle of radius 20$\lambda/D$, and more than 90\% inside a circle of radius 2$\lambda/D$). The expressions of $C_{L}$ (where L is the angular half-width covered by the mask) is written, in units of $\lambda/D$:

\begin{equation}
\begin{split}
C_{L}(\alpha,\beta) &= \int_{0}^{L} \cos \big( 2 \pi u \alpha \big) \sin \big( 2 \pi u \beta \big) \, du\\
&= \frac{\sin^{2}(\pi L (\alpha+\beta))}{2 \pi (\alpha+\beta)} - \frac{\sin^{2}(\pi L (\alpha-\beta))}{2\pi(\alpha-\beta)} \\
\end{split}
\label{C12}
\end{equation}

This concludes our analysis of the mathematical formalism used for the four-quadrant phase mask. Indeed, to compute the electric field $P(\tilde{x},\tilde{y})$ one only has to plug in the expressions of $C_{L}$ found in Eq.\ref{C12}. Note that in this case the mask cover a square region of the image plane centered on its origin.

A closed form for the Fourier transform of $P(\tilde{x},\tilde{y})$, and therefore the attenuated image of the star, could be found as long as no Lyot stop is introduced in that plane. As using a Lyot stop appears however mandatory, this additional step is not explored further.

%%%%

\subsection{Case of the vortex phase mask}

The VPM can be represented mathematically through the complex function $M_V(\theta)=e^{i \theta l}$, where $l$, the topological charge, is an even integer strictly greater than 0, and $\tan(\theta)=v/u$. The real part of this function is even (with respect to $u$ and $v$), while its imaginary part is odd. Contrary to the 4QPM, the phase induced by the VPM cannot be represented by the product of a function of $u$ and a function of $v$:
\begin{equation}
\begin{split}
M_{V}(\theta) & =\cos(\theta l) + i \sin(\theta l) \\
& = \cos^{l}(\theta) \, \sum_{k=0}^{l} i^{k} \, {l \choose k} \, \tan^{k}(\theta) \\
%& = \Big(\frac{u^2}{u^2+v^2}\Big)^{l} \, \sum_{k=0}^{2l} i^{k} \, {2l \choose k} \, (v/u)^{k} \\
\end{split}
\end{equation}
Because the phase induced by the mask depends on $\theta$, the integration in plane B is done with respect to the polar coordinates $\rho$ and $\theta$ instead of $u$ and $v$:
\begin{equation}
\begin{split}
P(\tilde{x},\tilde{y}) = i \lambda F e^{2 i \pi F/\lambda} & \int_{0}^{\infty}\int_{0}^{2\pi} E(\rho \cos(\theta),\rho \sin(\theta)) \\ 
& \times e^{i \theta l} \times e^{2i\pi \rho (\tilde{x} \cos(\theta) + \tilde{y} \sin(\theta))}\, \rho \, d\rho \, d\theta
\end{split}
\end{equation}
Plugging in the expression for the electric field $E(u,v)$ as a function of $A(x,y)$ gives:
\begin{equation}
\begin{split}
P(\tilde{x},\tilde{y}) = -e^{4 i \pi F/\lambda} \iint_{-D/2}^{D/2} A(x,y) \int_{0}^{\infty} \int_{0}^{2\pi} & e^{-2 i \pi \rho ((x-\tilde{x}) \cos(\theta)+(y-\tilde{y}) \sin(\theta))} \\
& \times e^{i \theta l} \, \rho \, d\rho \, d\theta \, dx \, dy \\
= -e^{4 i \pi F/\lambda} \iint_{-D/2}^{D/2} A(x,y) \times F(x,\tilde{x}, & y,\tilde{y}) \, dx \, dy
%= -16 e^{4 i \pi F/\lambda} & \iint_{0}^{\infty} A(x,y) \, F(x,\tilde{x},y,\tilde{y}) \, dx\, dy \\
\end{split}
\label{P2PV}
\end{equation}
Where $F(x,\tilde{x}, y,\tilde{y})$ denotes the double integral with respect to $\rho$ and $\theta$. One can notice that Eq. \ref{P2PV} can also be written as the convolution of $A(x,y)$ and a kernel $K(x,y)$ which is the Fourier transform of the mask:
\begin{equation}
\begin{split}
K(x,y)=\int_{0}^{\infty} \int_{0}^{2\pi} \rho \, e^{i \theta l} e^{-2 i \pi \rho (x \cos(\theta)+y \sin(\theta))}  \, d\rho \, d\theta
\end{split}
\label{ConvOne}
\end{equation}
In any case, the integral with respect to $\rho$ has a closed form if the domain of integration is finite and not semi-infinite:
\begin{equation}
\begin{split}
\Psi_{L}(a)=\int_{0}^{L} \rho \, e^{-2 i \pi \rho a} d\rho = \frac{(1 + 2 i \pi a L) \, e^{-2 i \pi a L}  -1}{4 \pi^2 a^2}
\end{split}
\end{equation}
A closed form can then be sought for the integral with respect to $\theta$. We only consider here the case of the topological charge $l=2$:
\begin{equation}
\begin{split}
& F(x,\tilde{x},y,\tilde{y}) = \int_{0}^{2\pi} e^{2i\theta} \, \Psi_{L}\Big( (x-\tilde{x}) \cos(\theta) + (y-\tilde{y}) \sin(\theta) \Big) \, d\theta \\
& = e^{2 i \phi(x,\tilde{x},y,\tilde{y})} \, \Big( \frac{J_{0}(2 \pi L \, r(x,\tilde{x},y,\tilde{y}) )-1}{\pi \, r(x,\tilde{x},y,\tilde{y})^2}+\frac{L \, J_{1}(2 \pi L \, r(x,\tilde{x},y,\tilde{y}) )}{r(x,\tilde{x},y,\tilde{y})} \Big)\\
& \textrm{where } r(x,\tilde{x},y,\tilde{y}) = \sqrt{(x-\tilde{x})^2+(y-\tilde{y})^2} \\
& \textrm{ and } \phi(x,\tilde{x},y,\tilde{y}) = \tan^{-1}\Big(\frac{y-\tilde{y}}{x-\tilde{x}}\Big)
\end{split}
\label{FinF}
\end{equation}
The electric field $P(\tilde{x},\tilde{y})$ can be computed using Eq.\ref{P2PV}, the expression of $F(x,\tilde{x},y,\tilde{y})$ being given in Eq.\ref{FinF}. Note that contrary to the case of the four-quadrant phase mask, and contrary to the case of a classical Fourier transform, the integrals with respect to $x$ and $y$ cannot be separated. This is a major obstacle to a fast computation of the electric field in plane C.
The symmetries of the apodization pattern $A(x,y)$ and of the real and imaginary parts of $F(x,\tilde{x},y,\tilde{y})$ can however be used to integrate over a smaller domain of plane A:
\begin{equation}
\begin{split}
& P(\tilde{x},\tilde{y}) = -\frac{1}{4}e^{4 i \pi F/\lambda} \iint_{0}^{D/2} A(x,y) \, \Big( F(x,\tilde{x}, y,\tilde{y}) + F(x,-\tilde{x}, y,-\tilde{y}) \\
& + e^{-4i\phi(x,\tilde{x}, y,-\tilde{y})} F(x,\tilde{x}, y,-\tilde{y}) + e^{-4i\phi(x,-\tilde{x}, y,\tilde{y})} F(x,-\tilde{x}, y,\tilde{y}) \Big) \, dx \, dy
\end{split}
\label{P2PVsmall}
\end{equation}
While the integration is then done over a fourth of the pupil, the function $F(x,\tilde{x}, y,\tilde{y})$ has to be evaluated four times if no simplification can be found and the total number of operation remains the same as in the previous case. The number of constraints though is reduced by four.

%%%%%%%%%%%%%%%%

\section{Discretization \& Optimization}
\label{DiscretizationOptimization}

To be suitable for computations and numerical optimizations, the equations derived previously must be converted to a discrete form. 

\subsection{Discretization}
\label{Discretization}
The axes in plane A are sampled with $N_{1}$ points, and with $N_{2}$ points in plane C. In the case of the 4QPM, and looking back at Eq.\ref{P2P}, this leads to writing that:
\begin{equation}
\begin{split}
P(\tilde{x}_{k},\tilde{y}_{l}) & = 16 \,e^{4 i \pi F / \lambda} \sum_{i=1}^{N_{1}} \sum_{j=1}^{N_{1}} A(x_{i},y_{j}) C_{1,L}(x_{i},\tilde{x}_{k}) \, C_{2,L}(y_{j},\tilde{y}_{l}) \, \Delta x  \, \Delta y \\
& = 4 \frac{D^2}{N_{1}^2} \,e^{4 i \pi F / \lambda} \sum_{i=1}^{N_{1}} \sum_{j=1}^{N_{1}} A(x_{i},y_{j}) C_{1,L}(x_{i},\tilde{x}_{k}) \, C_{2,L}(y_{j},\tilde{y}_{l}) \\
\textrm{with } x_{i} & =(i-\frac{1}{2}) \, \frac{D}{2N_{1}} \textrm{, } y_{j}=(j-\frac{1}{2}) \, \frac{D}{2N_{1}} \textrm{, } \\
\textrm{and } \tilde{x}_{k} & =(k-\frac{1}{2}) \, \frac{D}{2N_{2}} \textrm{, } \tilde{y}_{l}=(l-\frac{1}{2}) \, \frac{D}{2N_{2}} .
\end{split}
\label{DiscreteForm}
\end{equation}
Assuming a uniform distribution of points along both the x and y axes, $\Delta x$ and $\Delta y$ are replaced with $\frac{D}{2 N_{1}}$. Computing the value of the electric field for the $N_{2}^2$ points in plane C is done with a total of $N_{1}^2 N_{2}^2$ operations.
This number can be reduced by proceeding to this integration in two successive steps:
\begin{equation}
\begin{split}
& P_{temp}(\tilde{x}_{k},y_{j}) = 8 \frac{D}{N_{1}} \,e^{4 i \pi F / \lambda} \sum_{i=1}^{N_{1}} A(x_{i},y_{j}) \, C_{1,L}(x_{i},\tilde{x}_{k}) \\
& P(\tilde{x}_{k},\tilde{y}_{l}) = \frac{D}{2N_{1}} \,e^{4 i \pi F / \lambda} \sum_{j=1}^{N_{1}} P_{temp}(\tilde{x}_{k},y_{j}) \, C_{2,L}(y_{j},\tilde{y}_{l}) \\
\end{split}
\label{TwoSteps}
\end{equation}
And the complexity becomes $N_{1}^2 N_{2} + N_{2}^2 N_{1}$.

In the case of the VPM, using Eq.\ref{P2PV}, the integration is slightly different:
\begin{equation}
\begin{split}
P(\tilde{x}_{k},\tilde{y}_{l}) & = \frac{D^2}{4 N_{1}^2} \,e^{4 i \pi F / \lambda} \sum_{i=-N_{1}+1}^{N_{1}} \sum_{j=-N_{1}+1}^{N_{1}} A(x_{i},y_{j}) F(x_{i},\tilde{x}_{k},y_{j},\tilde{y}_{l}) \\
\end{split}
\label{DiscreteFormV}
\end{equation}
Using Eq.\ref{P2PVsmall} instead of Eq.\ref{P2PV} does not change the complexity, which equals $2 \times 16 \times N_{1}^2 N_{2}^2$ in both cases. Indeed, $F(x,\tilde{x},y,\tilde{y})$ is a complex function, and the real and imaginary parts must be considered separately in the optimization, which increases the number of computations by a factor two. Furthermore the entire pupil plane must be considered here, and four times as many points are required. It was previously showed that the parity properties of $F$ can be used to compute the field using only one quadrant of the pupil, however the actual number of computations remains the same. Finally, separating the double integral in two successive integrals actually increases the complexity to $2 \times (16 \times N_{1}^2 N_{2}^2 + 8 \times N_{2}^2 N_{1})$ (in this case $P_{temp}$ would be a function of 3 variables, and not two).

One may wonder if it would not be more efficient to compute the electric field in plane C through two successive Fourier transforms. The total complexity would go as low as $N_{1}^2 M + N_{1} M^2 + 2\times(M^2 N_{2} + M N_{2}^2)$, where M would be the number of points used to sample the electric field along both axes of plane B. This complexity assumes a 2-axes symmetry in the entrance pupil plane, and relies on the physics of the VPM, so that the electric field can be computed in only one quadrant of planes A, B and C. The 2-axes symmetry is a property shared by many telescopes such as the VLT, the Subaru telescope and the Gemini telescope. Thanks to the properties of the Fourier transform, the same 2-axes symmetry is found in the electric field in plane B, and the VPM then creates in plane C an electric field whose real part presents the same symmetry, and an imaginary part that presents a 2-axes anti-symmetry. Hence, it is possible to use these properties to compute the electric field in only one quadrant of each planes, thus reducing the number of computations by a factor 4 in each transform.

%The cases where the electric field in plane B is explicitly computed or not must also be considered separately.

Realistic values for $N_{1}, N_{2}$ and $M$ must be chosen to assess the interest of this method over the previous one. If the electric field is not computed explicitly, only the pupil planes must be correctly sampled. Both the 4QPM and the VPM, used with a circular clear aperture, produce ripples in the reimaged pupil plane with a periodicity that is the invert of the angular extension of the focal phase mask. Hence, if the mask extends up to L $\lambda/D$, 2 L ripples appear in plane C. Assuming that the patterns that the apodizers produce present the same periodicity, the value of $N_1$ and $N_2$ must be chosen accordingly, i.e. they must be large enough to correctly sample this signal. To obey the Nyquist-Shannon sampling theorem, each period should be sampled with 2 points, which gives $N_2 = 2 \times L$ (only one quadrant of the pupil plane is actually sampled). Thus, observing up to 32 $\lambda/D$ implies that $N_2$ must at least equals 64. It might be preferable to consider a higher number $N_{1}$ of points in plane A since, when optimizing shaped pupils in two dimensions, the quasi-binary transmissions that are found can be rounded without changing the PSF properties as long as $N_1$ is large enough (a few hundred points for contrast as low as $10^{-7}$ and a thousand points for contrast values as low as $10^{-9}$). While algorithms exist to turn a gray apodization pattern into a binary one - as it is for example done when computing microdots apodizers \citep{Martinez2009} - a native binary solution would have the advantage of preventing any potential impact on the performance of the mask that the non-binary to binary conversion process would have.

In the case where $N_{1} = 128$  and $N_{2} = 64$, the complexity is $1.6 \times 10^{6}$ for the 4QPM and $1.3 \times 10^{8}$ for the VPM. Choosing $N_{1} = 512$ and $N_{2}= 64$ makes it $1.9 \times 10^{7}$ for the 4QPM and $2.1 \times 10^{9}$ for the VPM.

If the electric field in plane B is explicitly sampled, it is important to sample each $\lambda/D$ angular distance unit with a high enough number of points, as errors due to a poor sampling in plane B will propagate to plane C. Although the Nyquist-Shannon sampling theorem requires a minimum of 2 points per $\lambda/D$, the specificities of the VVC demand a higher number of points. It is also likely that the higher the topological charge, the higher the number of points since the phase variation across the image plane becomes steeper. In fact, in the case of a clear aperture, a minimum number of 4 points per $\lambda/D$ is necessary to obtain contrast values down to $10^{-8}$ when the charge is 2. When the charge is twice as high, 8 points are necessary to recover the same contrast.

If $N_{1} = 128$  and $N_{2} = 64$, and the same angular extension of 32 $\lambda/D$ is chosen in plane B, then the complexity of this computation equals $7.3 \times 10^6$ if 4 points are used to sample one $\lambda/D$, and it becomes $2.3 \times 10^{7}$ if 8 points are chosen instead. These numbers are respectively 4.6 and 14.4 as high as what was required with the 4QPM, but also about 300 and 100 \emph{less} than what would be required for the VVC if the electric field in plane C was computed directly, using Eq.\ref{DiscreteFormV}.

\subsection{Optimization model}
\label{OptimizationModel}

The optimization of the pupil mask is very similar to the optimization of 2-dimensions shaped pupils, as described in \cite{Carlotti2009} and \cite{Vanderbei2012}. Though in both cases the problems are convex linear, the constraints that are specified here are not computed using Fourier transforms, but through the transforms described in Eq.\ref{P2P} and \ref{P2PV}.

The problem consists in maximizing the total transmission of the mask: \\$\sum_{i=1}^{N} \sum_{j=1}^{N} A(x_{i},y_{j}) \Delta x \Delta y$ under the following constraints set on the values of the transmission of the apodizer, and on the amplitude of the electric field in the reimaged pupil plane, or plane C:
\begin{equation}
\begin{split}
 0 < A(x_{i},y_{j}) < 1 &  \textrm{   , with    } \{ x_{i}, y_{j} \} \in \Delta_{A} \\
-10^{-c/2} \le P(\tilde{x}_{k},\tilde{y}_{l}) \le 10^{-c/2} & \textrm{   , with    } \{ \tilde{x}_{k}, \tilde{y}_{l} \} \in \Delta_{C}
\end{split}
\end{equation}
where $\Delta_{A}$ is the region defined by the telescope's aperture, $\Delta_{C}$ is the region defined by the Lyot stop, and $c$ measures the attenuation of the intensity in the reimaged pupil plane in a logarithmic scale. With a clear aperture, this attenuation factor is as small as $10^{-5}$ in the center of the pupil, although this number depends on the size of the phase mask, and becomes increasingly smaller as the mask becomes larger. Nonetheless, it is a good starting point for the optimization problems that must be solved.

The optimization model is coded using the AMPL language, and it is solved using the LOQO solver developed by Robert Vanderbei (see \cite{Vanderbei1998}).

\section{Results}
\label{Simulations}

The implementation of the two linear transforms introduced in Eq.\ref{P2P} and Eq.\ref{P2PV} makes it possible to study the effect of the central obstruction on the contrast in the final image plane D, as well the impact of smaller or larger mask size $L$. In the case of the 4QPM it also allows numerical optimizations to be performed efficiently.

\subsection{Impact of the central obstruction}

\begin{figure}[]
\centering
\begin{tabular}{cc}
\subfigure[]{\includegraphics[width=0.5\columnwidth]{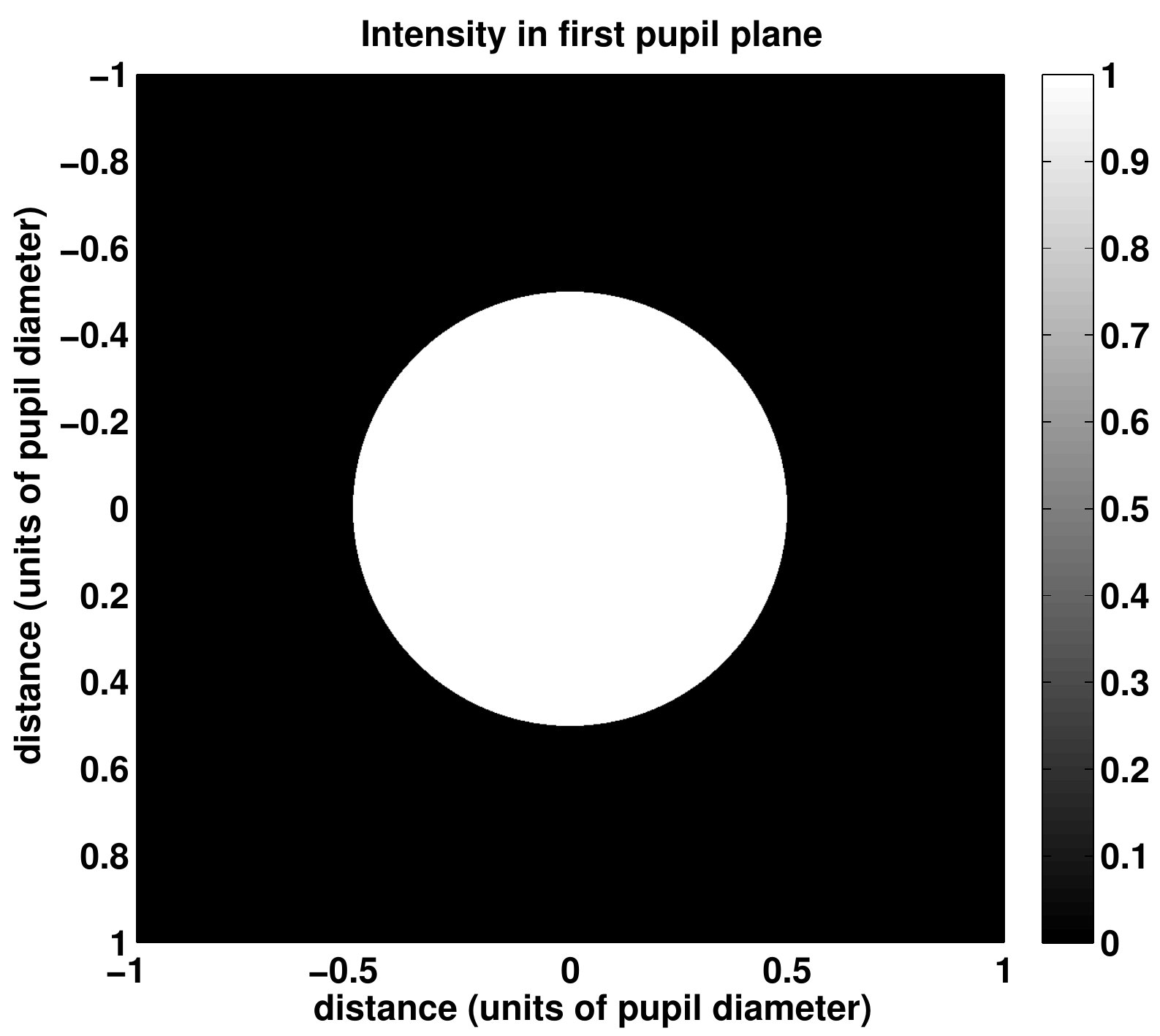}} \subfigure[]{\includegraphics[width=0.5\columnwidth]{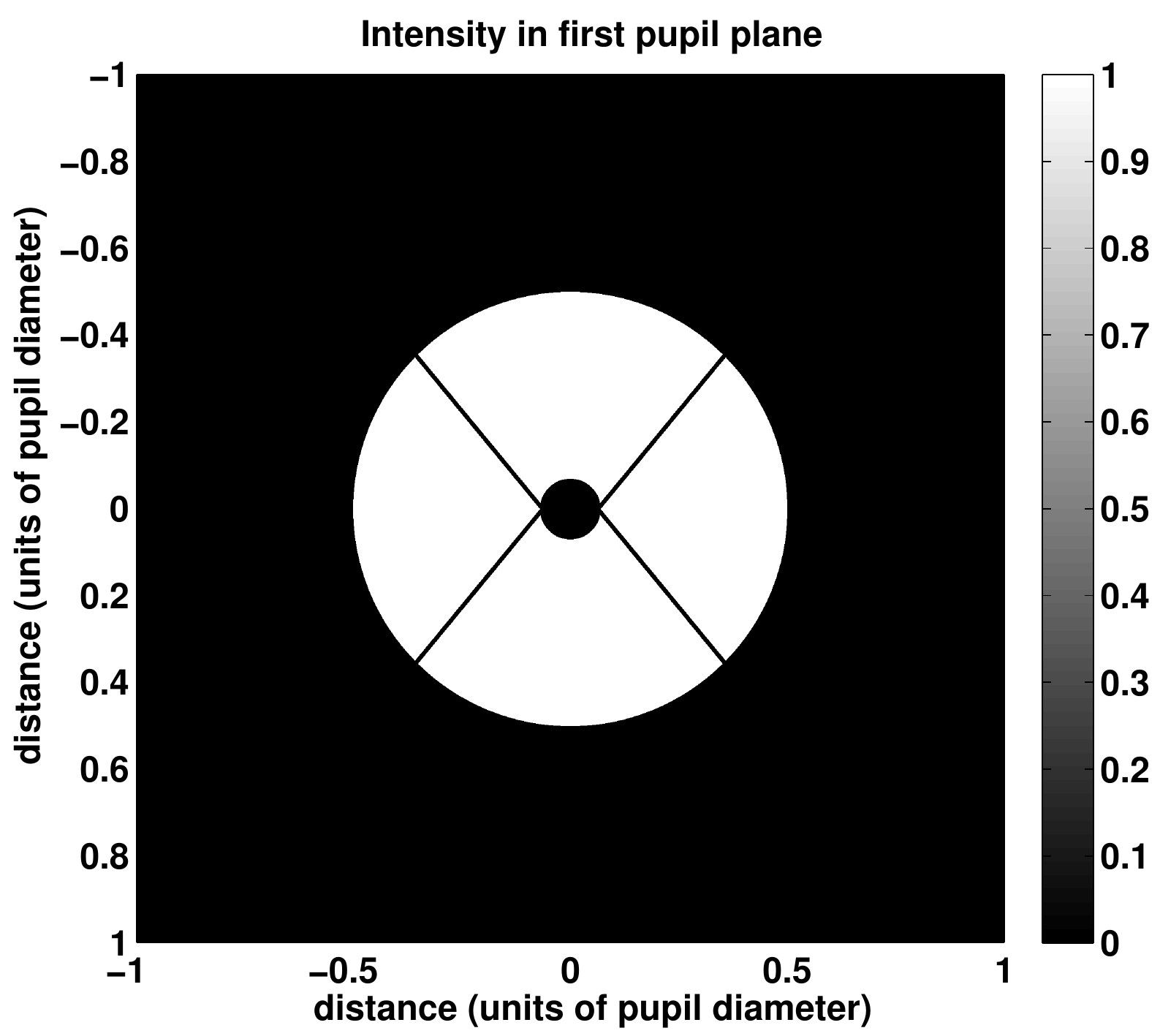}} \\
\subfigure[]{\includegraphics[width=0.5\columnwidth]{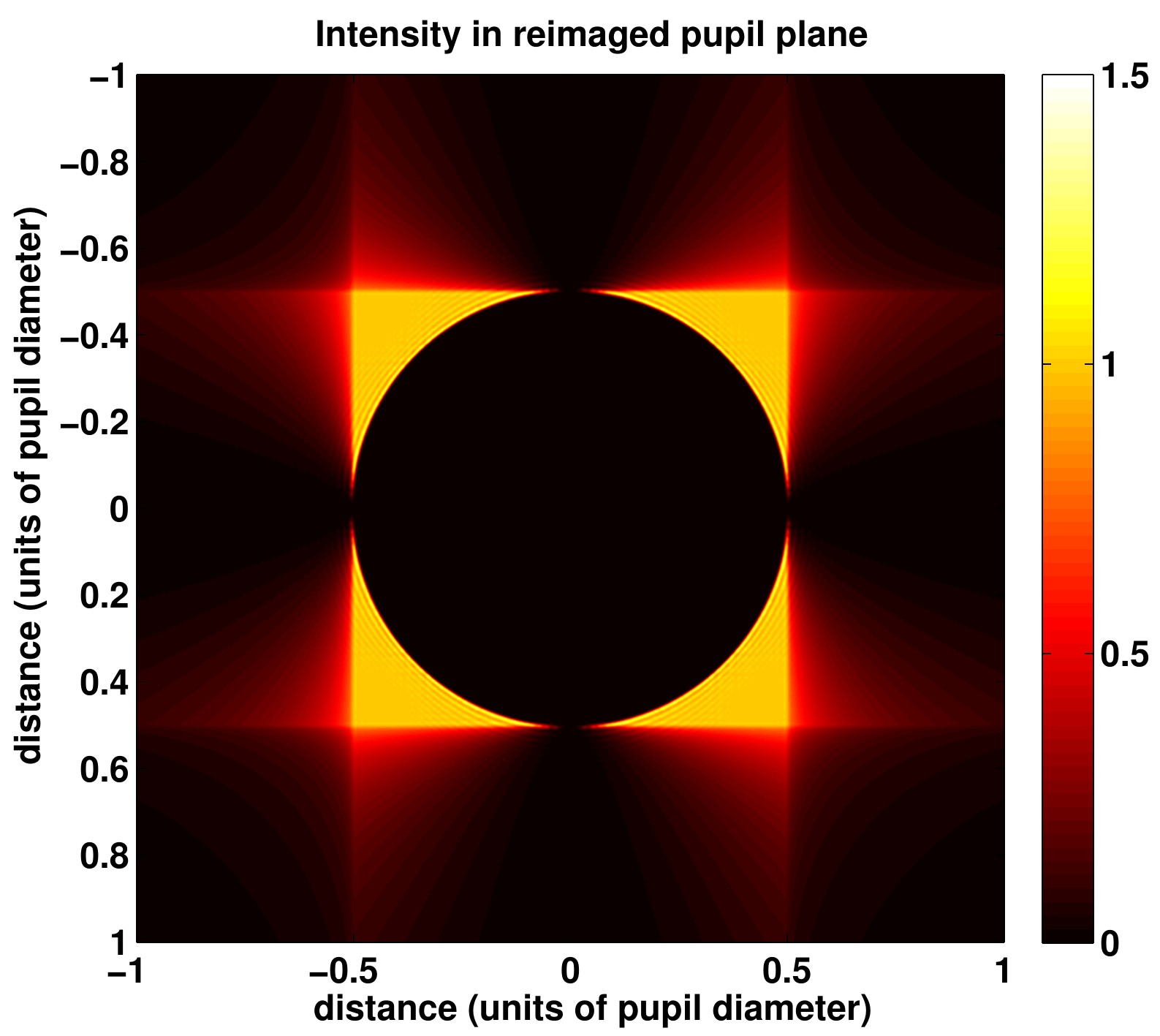}} \subfigure[]{\includegraphics[width=0.5\columnwidth]{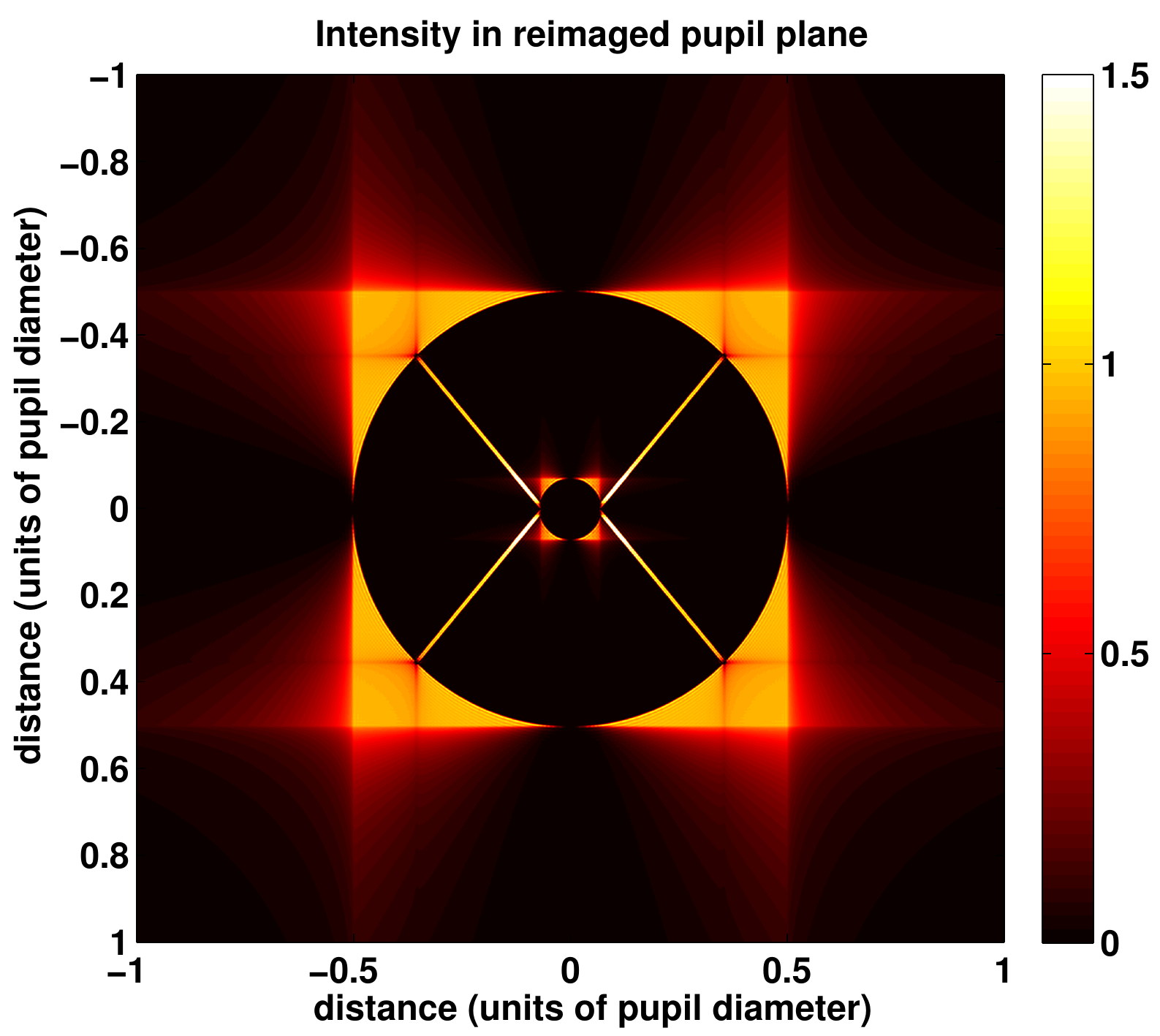}} \\
\subfigure[]{\includegraphics[width=0.5\columnwidth]{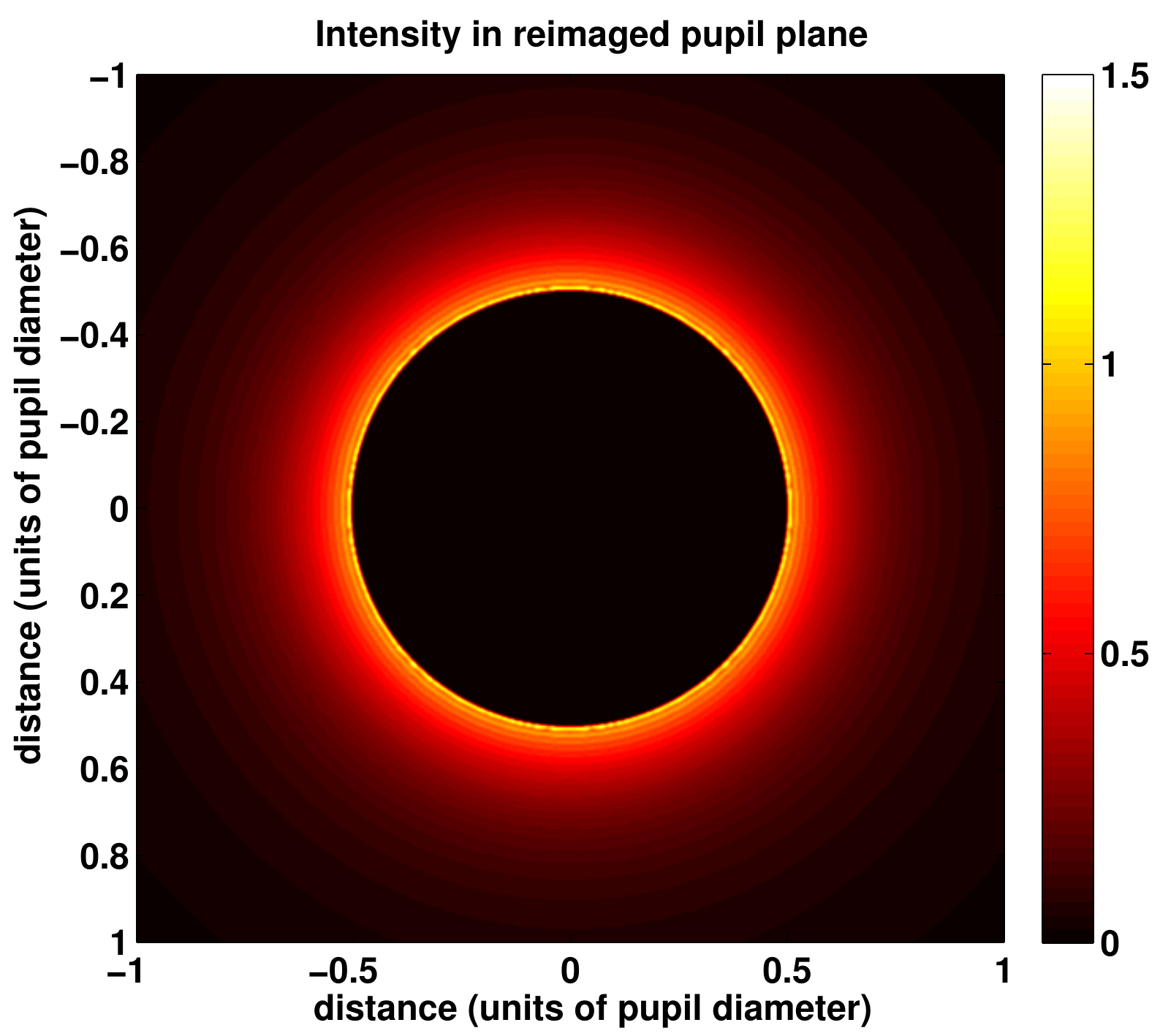}} \subfigure[]{\includegraphics[width=0.5\columnwidth]{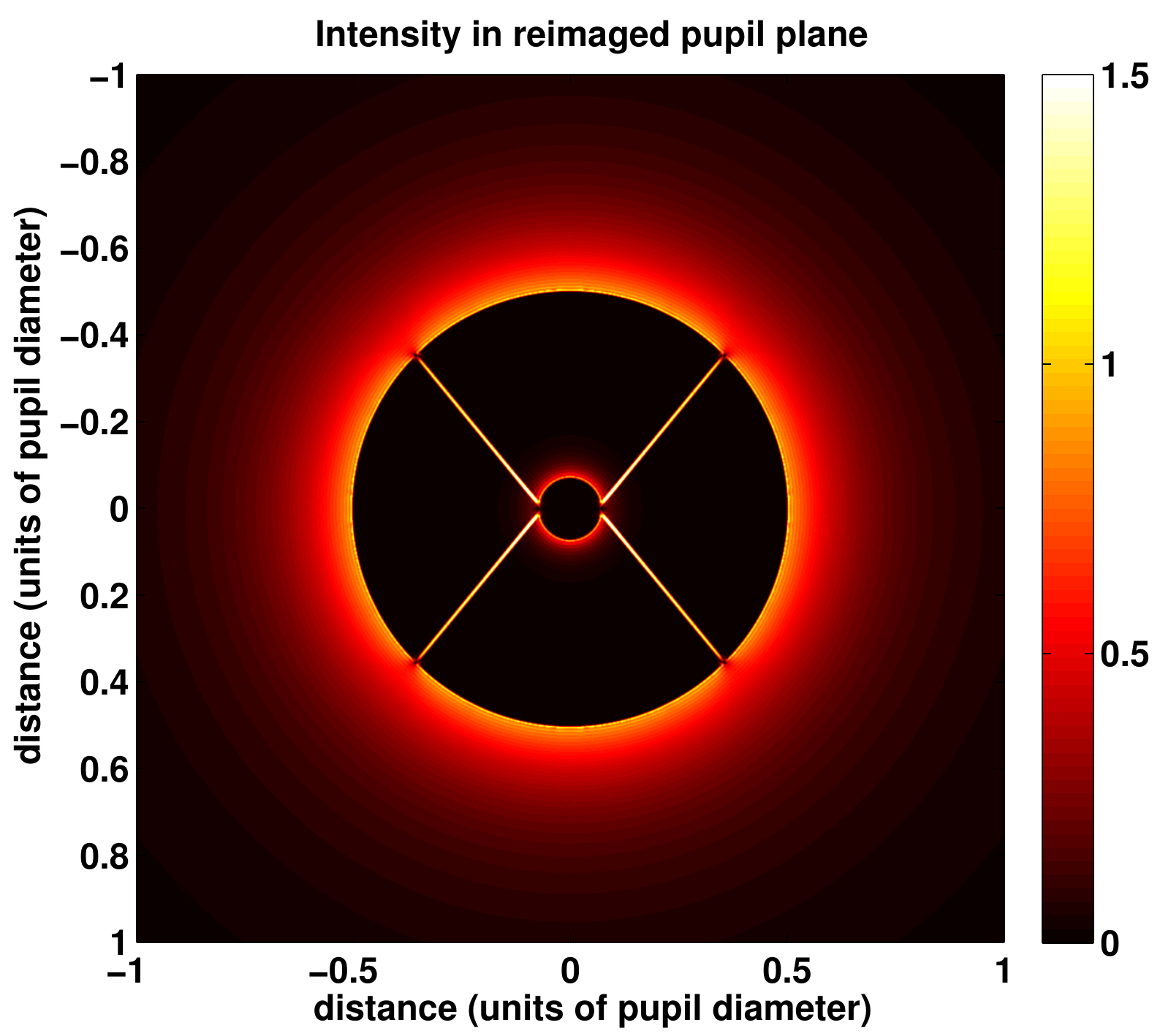}}
\end{tabular}
\caption{Fig. (a) \& (b): normalized intensity in plane A for a clear and and centrally obstructed apertures. Fig. (c) \& (d): normalized intensity in plane C for the same apertures, after a 4QPM is located in plane B, extending up to $128 \lambda/D$. Fig. (e) \& (f): normalized intensity in plane C for the same apertures, after a VPM is located in plane B (its angular extension is the same).}
\label{Reality}
\centering
\end{figure}

The transforms displayed in Eq.\ref{P2P} and Eq.\ref{P2PV} are here used in two basic cases: one is a circular clear aperture corresponding to an off-axis telescope, and the other is an on-axis VLT-like obstructed aperture (the details of the geometry were found in \cite{Guerri2008}). The normalized intensity that can be observed in the entrance pupil and the Lyot plane are displayed in Fig. \ref{Reality}. In both cases the linear number of points used to discretize the pupil in plane A equals 512, and the focal phase masks extend up to $128 \lambda/D$. In plane C, whereas all light is rejected outside the clear aperture, some light is diffracted inside the centrally obstructed aperture, reducing the attenuation of the on-axis light, as it can be seen in images (d) and (f) of Fig. \ref{Reality}.

\begin{figure}[]
\centering
\begin{tabular}{cc}
\subfigure[]{\includegraphics[width=0.45\columnwidth]{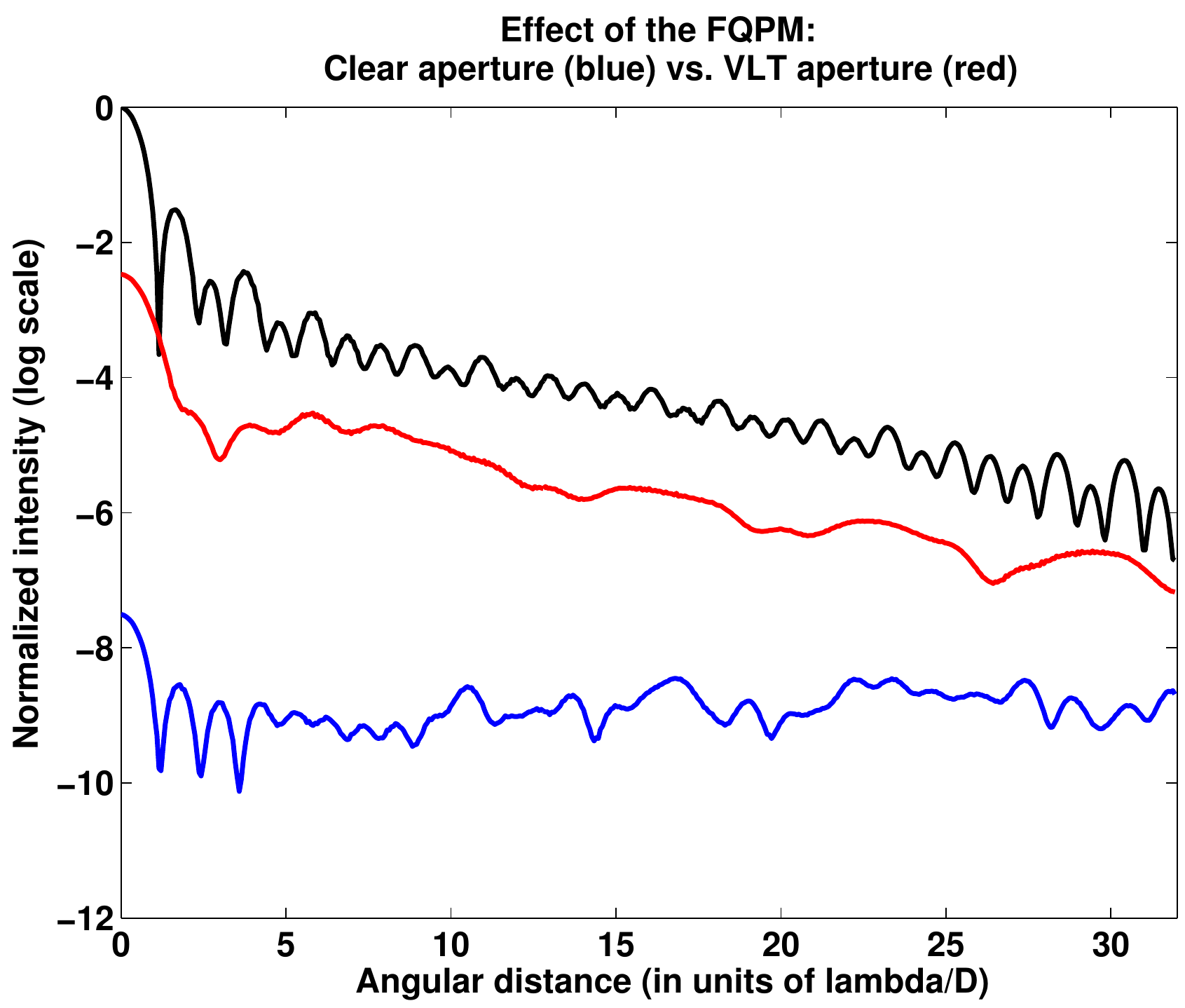}} \hspace{2mm} \subfigure[]{\includegraphics[width=0.45\columnwidth]{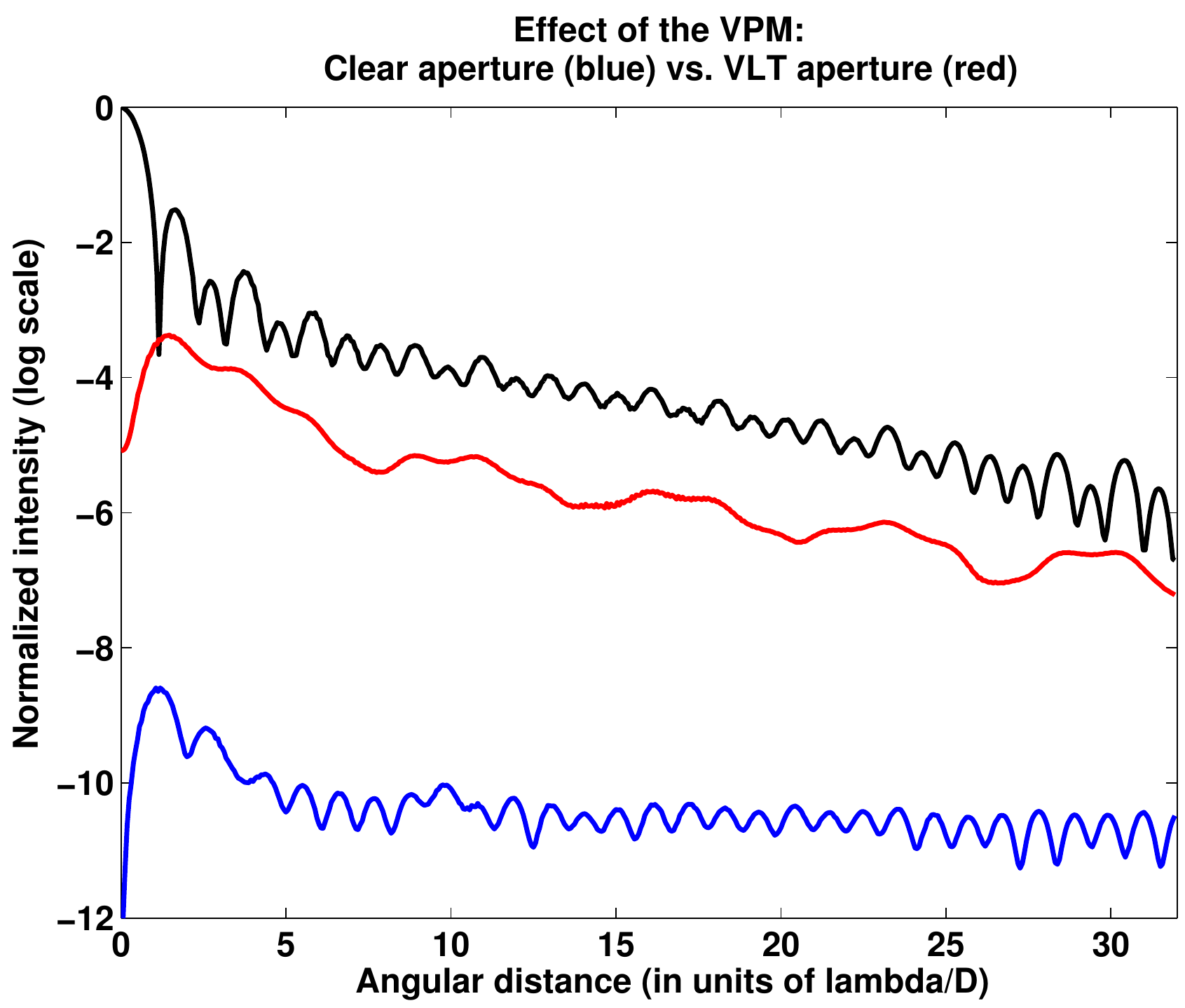}}
\end{tabular}
\caption{Azimuthal averages of the PSF of (a) the 4QPM coronagraph and (b) the VVC, for the two apertures shown in Fig.\ref{Reality}. The blue line corresponds to the case of a clear circular aperture, while the red line corresponds to the case of the VLT aperture. In both cases, a Lyot stop with a transmission of about 90\% was used to try and mitigate diffraction effects due to the finite size of the telescope aperture.}
\label{RealityPSF}
\centering
\end{figure}

The computation of the PSFs showed in Fig.\ref{RealityPSF} assumes the presence of a Lyot-stop in plane C. While two different Lyot stops must be used (one for the clear aperture, and one for the VLT-like aperture), both have a relative transmission of about 90\% with respect to the main aperture. The 4QPM used with the clear aperture cancels the starlight and creates a $10^{-9}$ contrast floor. The VPM performs a little better and creates a $10^{-10}$ contrast floor. Theoretically, both the 4QPM and the VPM cancels completely the on-axis light and this non-infinitely small value can be explained by (a) the finite size of the phase mask and (b) the necessary discretization of the pupil.

The performance of both phase masks with the centrally obstructed aperture of the VLT is significantly worse. Although the maximum intensity of the star is reduced by a factor $10^{2.4} \approx 250$, the coronagraphic PSF remains only at about one to two orders of magnitude below the telescope's PSF.

The contrast curves showed in Fig.\ref{RealityPSF} do not take into account the partial cancellation of the planet's light when it happens to be too close to the star. Following the definition proposed in \cite{Guyon2006}, the inner working angle (IWA) is defined as the distance for which no more than 50\% of the planet's light is cancelled. In the case of the 4QPM, this IWA is theoretically of about 1$\lambda/D$ but the planet's light is also attenuated if the planet is located on one of the phase transition axes of the mask, in which case the edge acts as a phase knife coronagraph canceling 80\% of the light.

\subsection{Impact of the finite angular size of the mask}
\label{ImpactSize}

In practice, whether the 4QPM or the VPM is used, the phase mask can always only cover a finite area of the image plane. In the rigorous analytical demonstration of the nulling property of the 4QPM, it is however assumed that the mask covers the entire image plane, and it is thus interesting to study the impact a finite angular size of the mask has on the PSF in the final image plane. The angular size $L$ of the mask is a parameter that must be set when computing the electric field in the Lyot plane, and Eq.\ref{P2P} can be conveniently used to assess the evolution of the contrast in plane D as function of $L$. 

\begin{figure}[]
\centering
\begin{tabular}{cc}
\subfigure[]{\includegraphics[width=0.45\columnwidth]{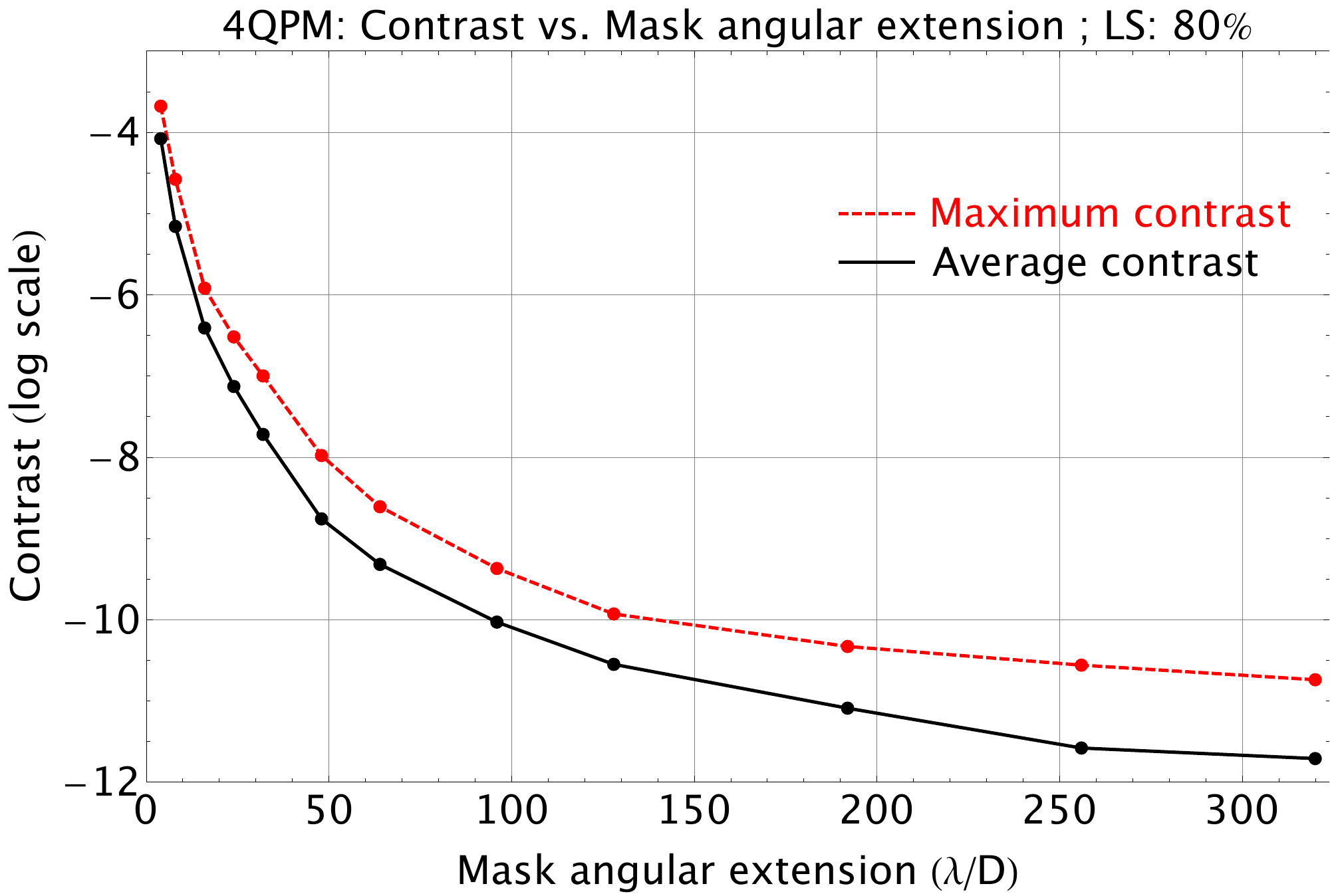}} \hspace{2mm} \subfigure[]{\includegraphics[width=0.45\columnwidth]{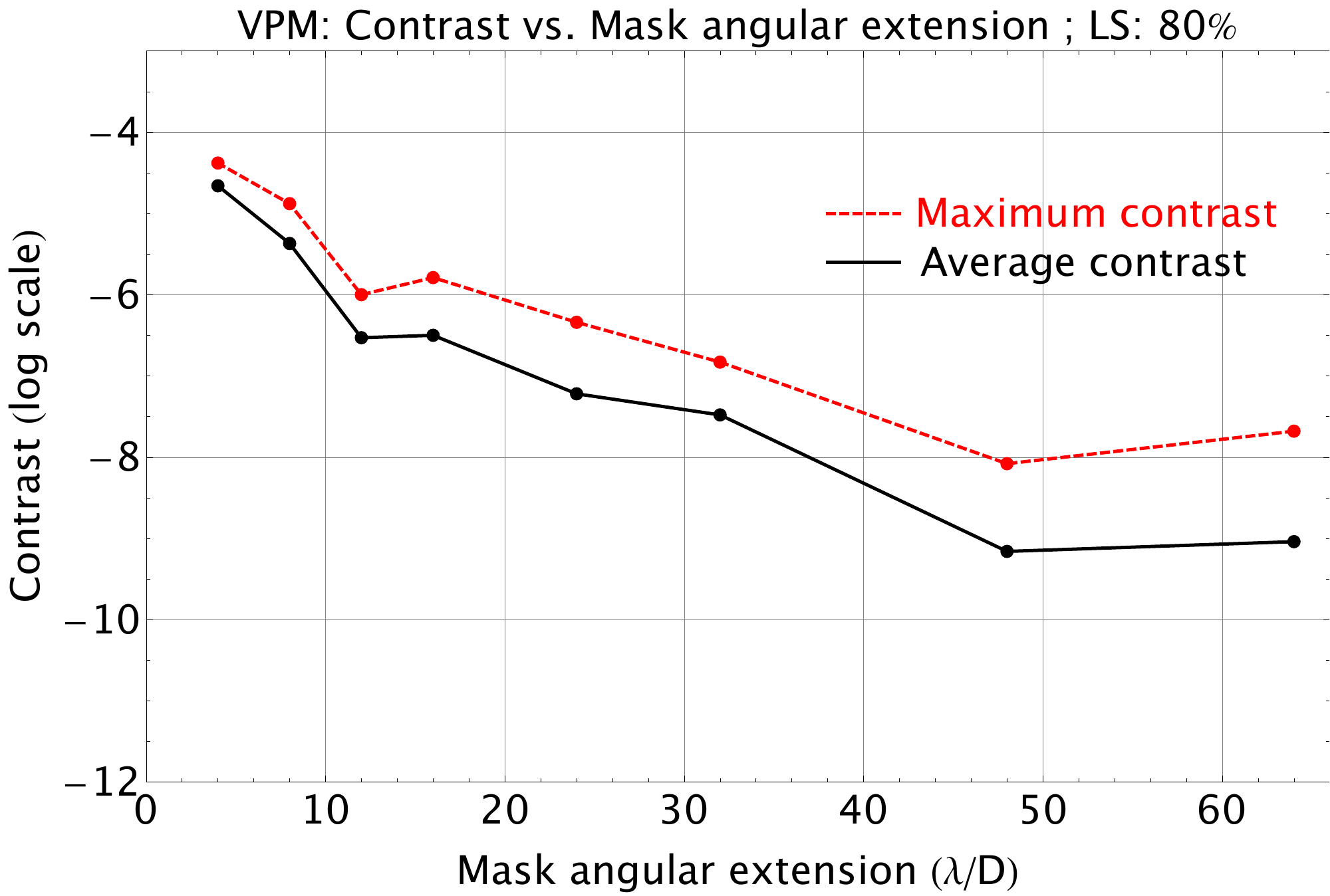}}
\end{tabular}
\caption{Evolution of the mean contrast (solid black line) and the maximum contrast (dashed red line) in plane D of a 4QPM coronagraph (left) and a VVC (right) used with a circular clear aperture, for a mask radius $L$ up to 320$\lambda/D$ (for the 4QPM) and up to 64$\lambda/D$ (for the VPM). An 80\% transmission Lyot stop was used in this case, and both pupil planes A and C were discretized over 4096 by 4096 points (for the 4QPM) and over 512 by 512 points (for the VPM).}
\label{FiniteMaskSize}
\end{figure}

As it is presented in Fig.\ref{FiniteMaskSize}, both the mean and the maximum contrasts decrease with higher values of $L$. In the case of the 4QPM, the diffraction effects caused by the finite size of the mask lead to a $10^{-5.9}$ maximum contrast for $L=16 \lambda/D$, a $10^{-8}$ maximum contrast for $L=48 \lambda/D$, and $10^{-10}$ maximum contrast for $L\approx 140 \lambda/D$. The case of the VPM is rather similar, although the decrease of the contrast value is less monotonous.

These values were computed assuming a clear circular aperture without a central obstruction, and without spiders. In both pupil planes A and C, the electric field was computed over 4096 by 4096 points in the case of the 4QPM, and over 512 by 512 points in the case of the VPM (this smaller number being chosen because of the higher complexity of the computation). The mean and maximum contrast values were computed from azimuthally averaged PSFs, over an angular range going from 0 to 48$\lambda/D$ for $L \le 64$, and from 0 to 0.75$L$ otherwise (this was done so as to mitigate diffraction effects occurring near the outer edge of the mask). The Lyot stop was chosen in this case to have an 80\% transmission. A more transmissive Lyot stop (90\% transmission) leads to similar but slower decrease, with a $10^{-10}$ maximum contrast being reached for $L=$190$\lambda/D$.

\subsection{Results of the optimizations}

\begin{figure}
\centering
\begin{tabular}{cc}
\subfigure[]{\includegraphics[width=0.45\columnwidth]{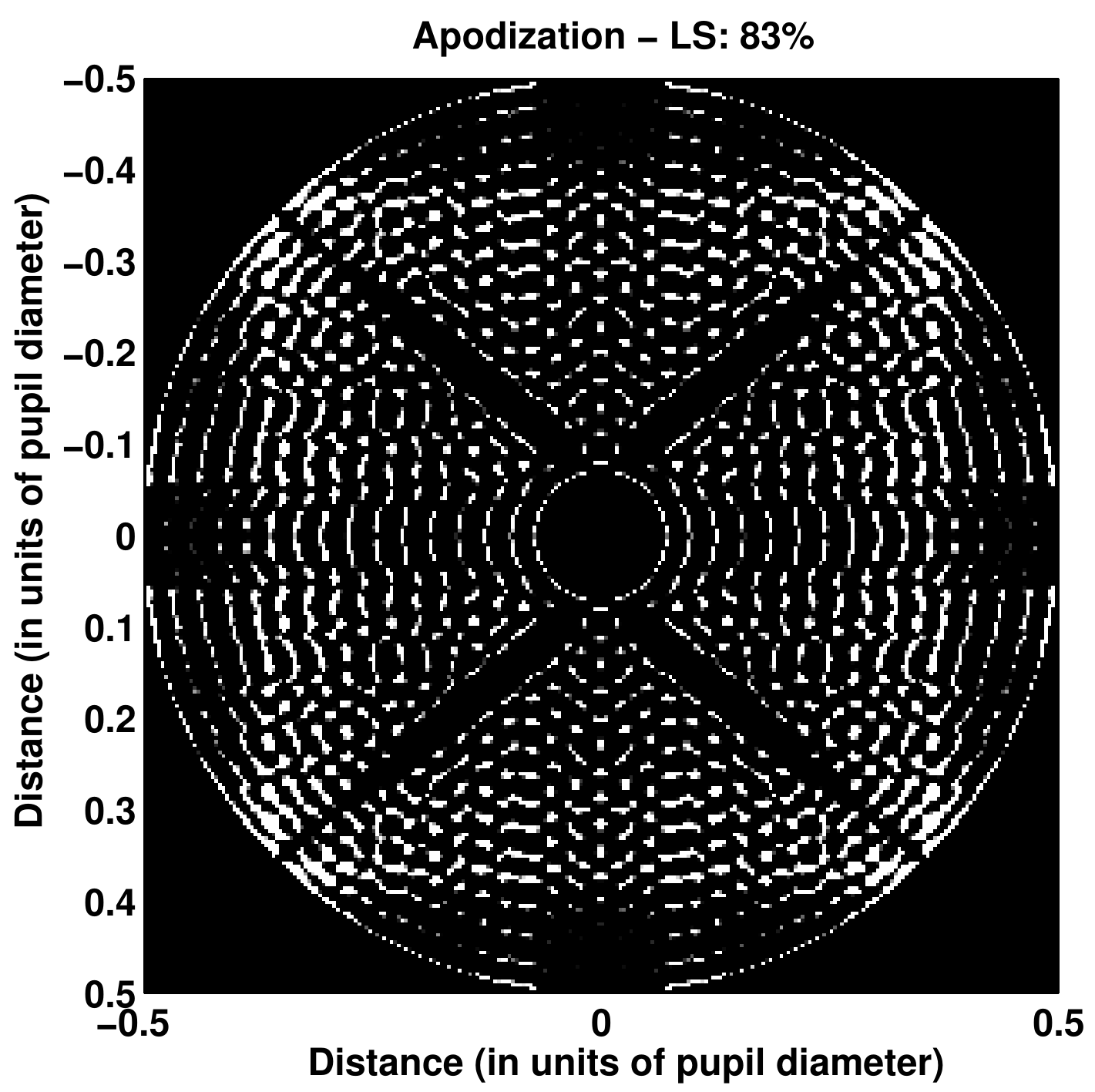}} \qquad \subfigure[]{\includegraphics[width=0.45\columnwidth]{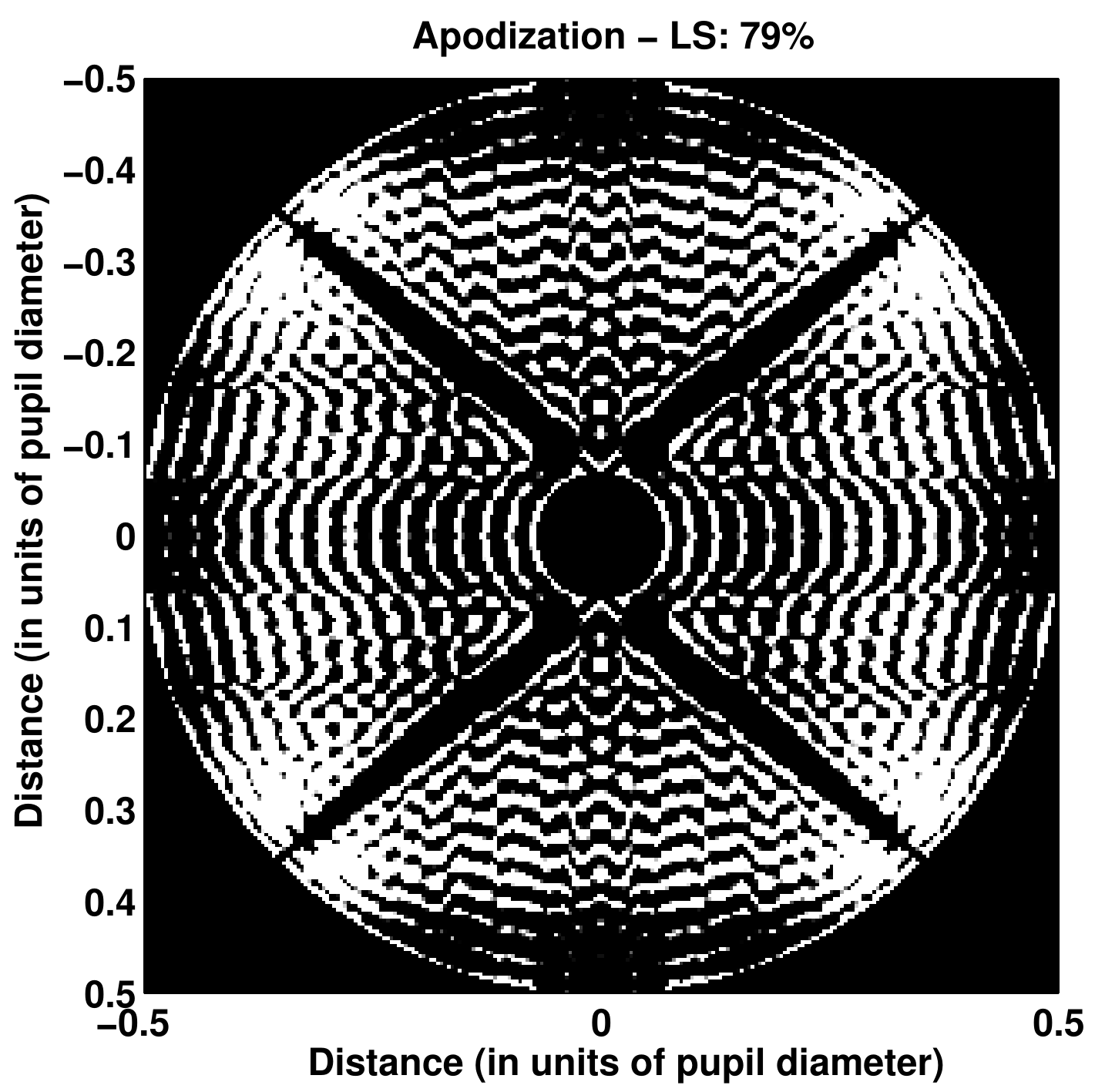}} \\
\subfigure[]{\includegraphics[width=0.45\columnwidth]{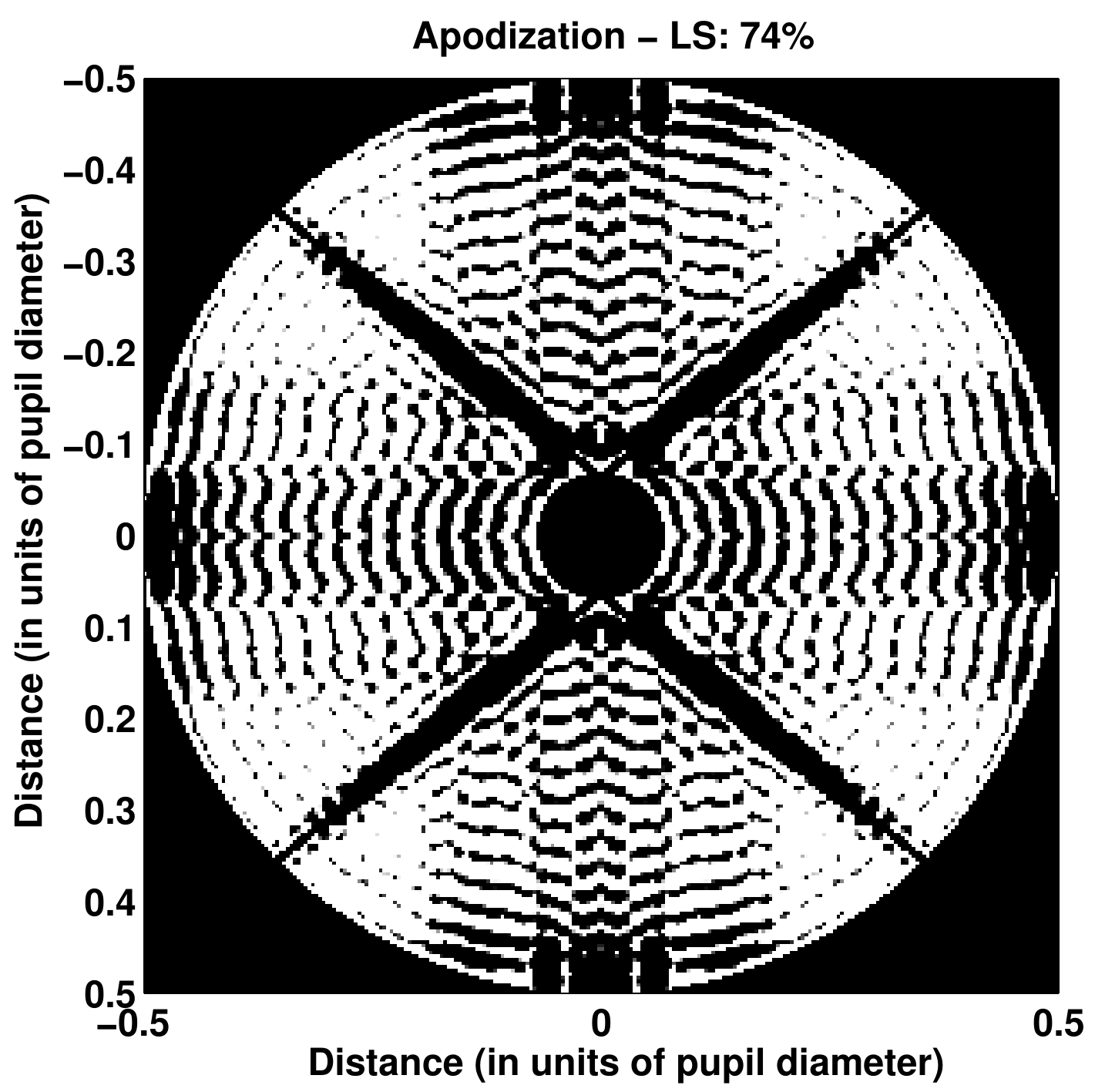}} \qquad \subfigure[]{\includegraphics[width=0.45\columnwidth]{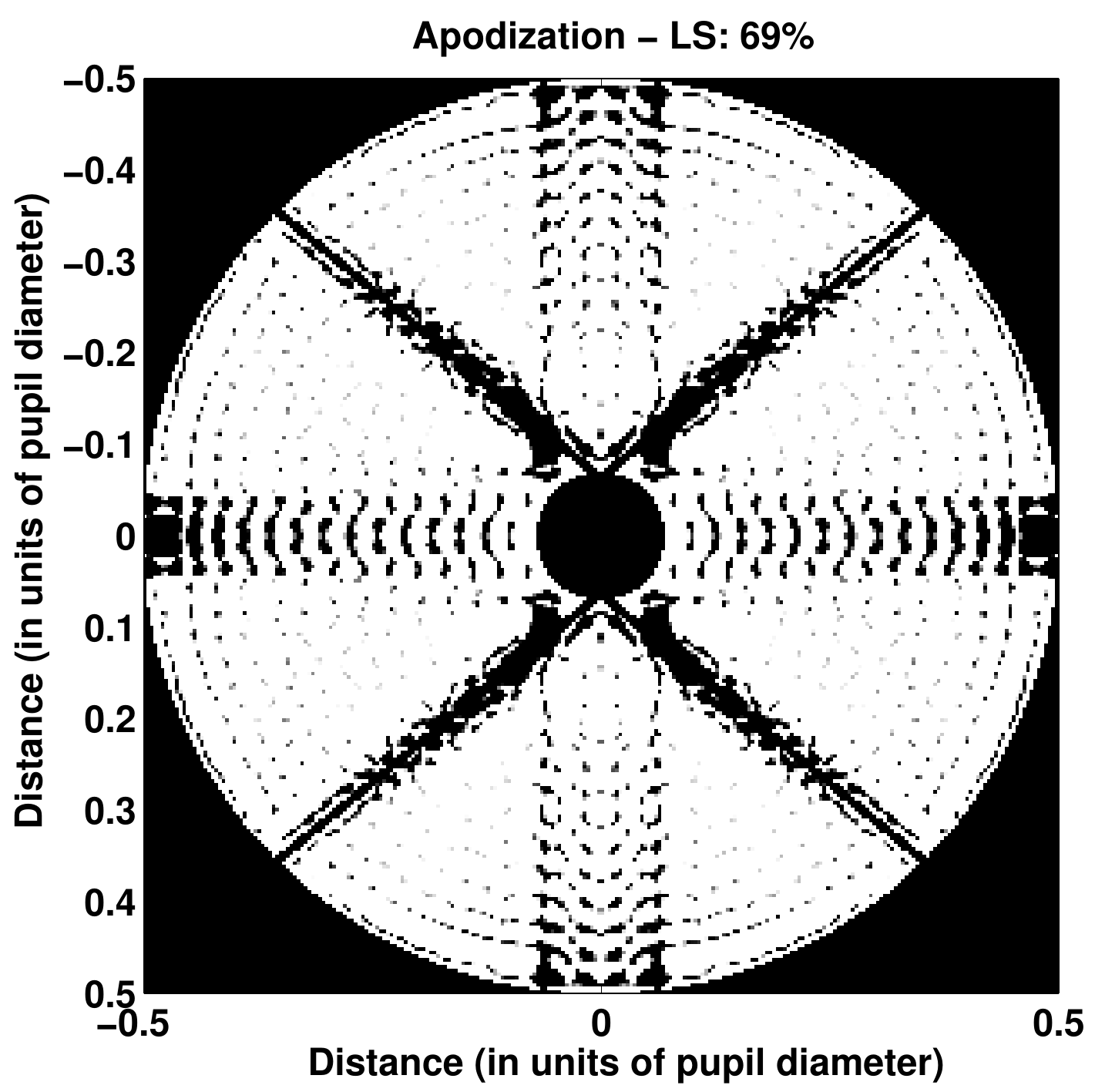}} \\
\subfigure[]{\includegraphics[width=0.45\columnwidth]{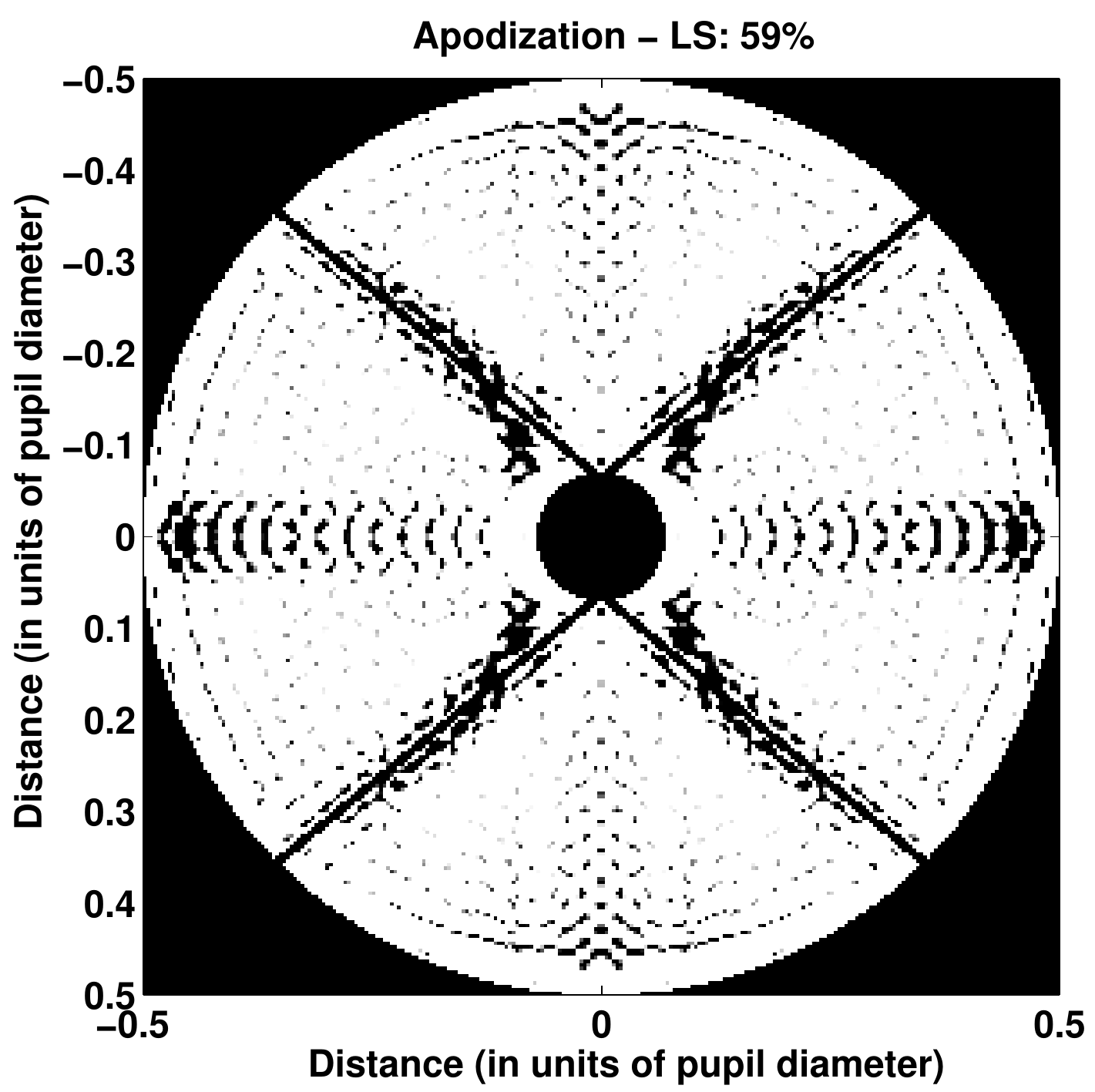}}
\end{tabular}
\caption{Transmitted intensities of the five apodizers optimized for the 4QPM at the VLT and a Lyot stop with a transmission of (a) 83\%, (b) 79\%, (c) 74\%, (d) 69\%, (e) 59\%. White denotes a unit transmission, and black a zero transmission.}
\label{Apodizers4Q}
\end{figure}

A 3.3 GHz processor out of a 24 cores machine, with a total RAM of 192 GB was used to solve the optimization problems. Since in practice this machine is shared with many users, only up to 8GB are available for one optimization process. Because of its lower complexity, the case of the 4QPM was considered first. Five different apodizers have been computed for this phase mask (see Fig.\ref{Apodizers4Q}) and for a VLT-like aperture. In the case of the VPM, the complexity of the optimization process is not much higher, but too much RAM was required and no results have yet been obtained for this mask.

\begin{figure}[]
\centerline{\includegraphics[width=0.8\columnwidth]{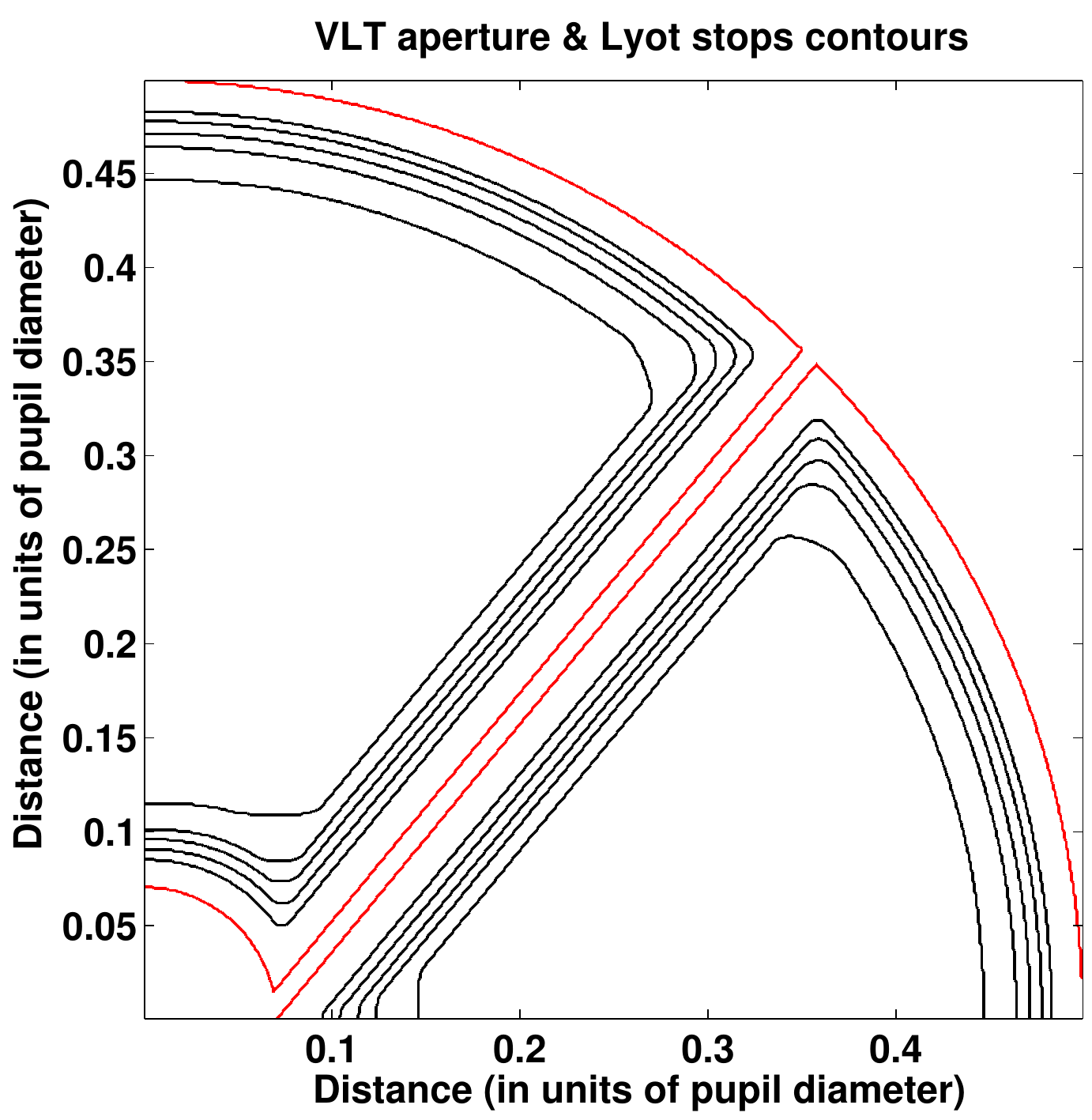}}
\caption{Contours of the VLT aperture (red) and of the five Lyot stops (black) with the respective transmissions of 83, 79, 74, 69 and 59\%.}
\label{LS}
\end{figure}

Every apodizer has been designed to attenuate the diffracted light inside the area defined by a specific Lyot stop (see Fig.\ref{LS}). The five Lyot stops were computed by oversizing the central obstruction and the spiders of the telescope, while undersizing its outer diameter. To do that, an image filtering process was used: an array describing the telescope aperture is convolved with unit matrix. The result is a blurred image of the aperture. Rounding it creates a Lyot stop. The transmission and resolution losses are adjusted by the relative size of the unit matrix and the aperture. An array of 1024 by 1024 points was used for the aperture to accurately change the transmission in the pupil plane. The Lyot stops were then interpolated in two dimensions to reduce their size. The transmissions $T_{LS}$ of the five Lyot stops range from 59\% to 83\%. These transmissions are defined as the ratio of the total intensity going through the masks with the total intensity going through the initial aperture of the telescope.

Indeed, each apodizer is optimized over an array of 128 by 128 pixels covering one quadrant of plane A. In every cases, the angular extension $L$ of the phase mask is 32$\lambda/D$. The electric field in plane C is also computed over one quadrant, and with the same number of points as in plane A. An average time of a day and a half were necessary to compute each apodizer.

\begin{figure}[]
\centering
\begin{tabular}{cc}
\subfigure[]{\includegraphics[width=0.5\columnwidth]{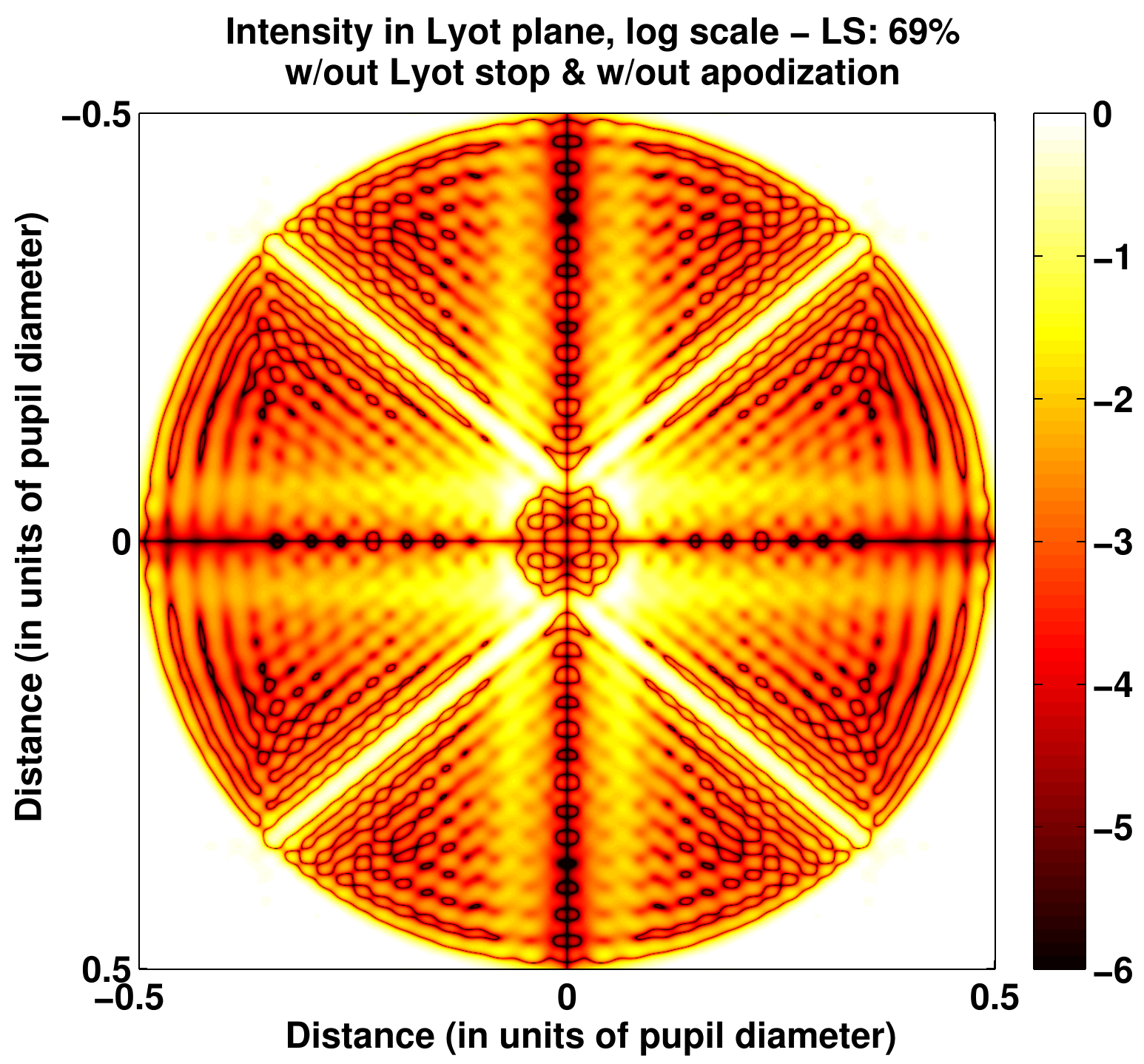}} \subfigure[]{\includegraphics[width=0.5\columnwidth]{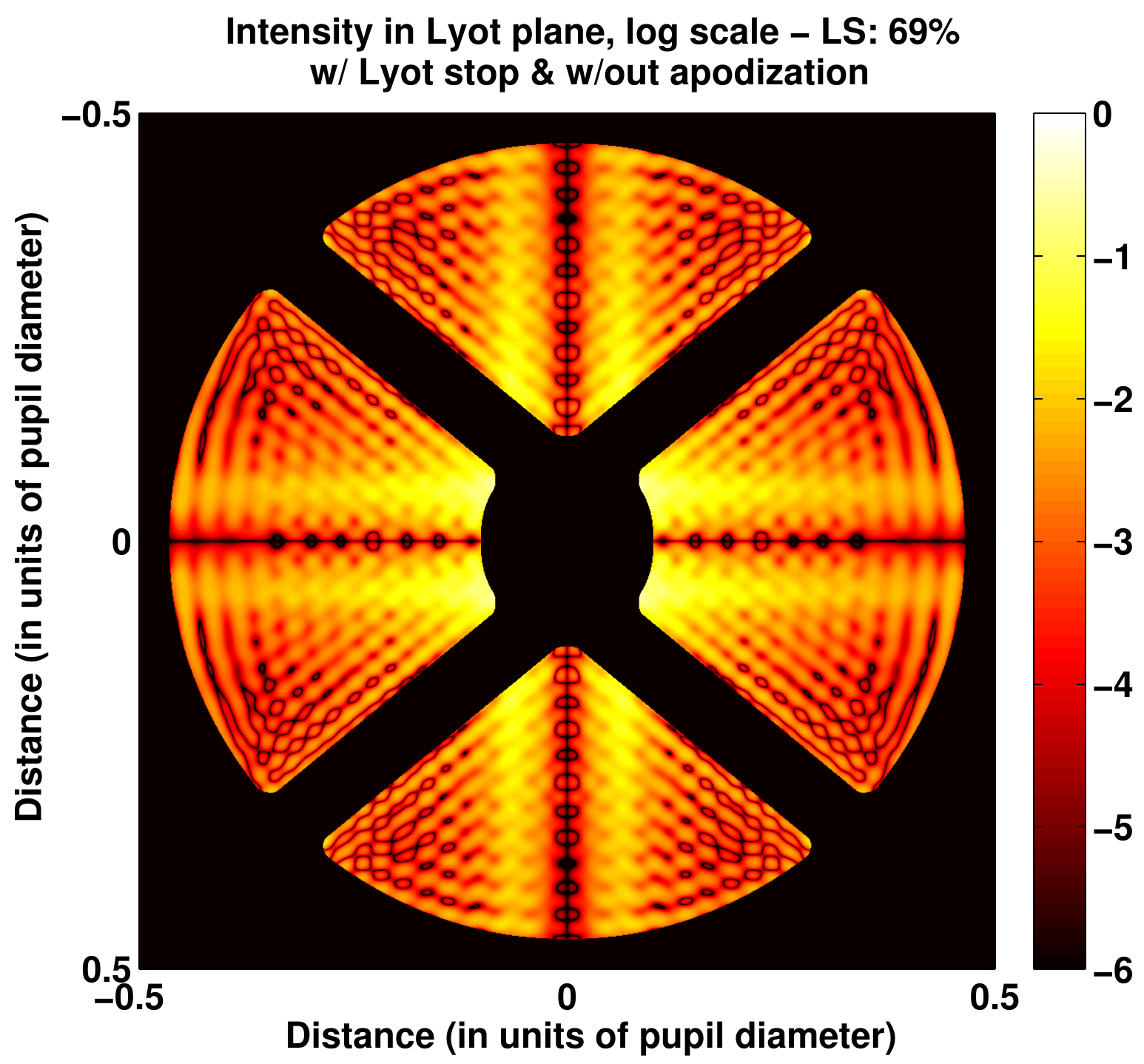}} \\
\subfigure[]{\includegraphics[width=0.5\columnwidth]{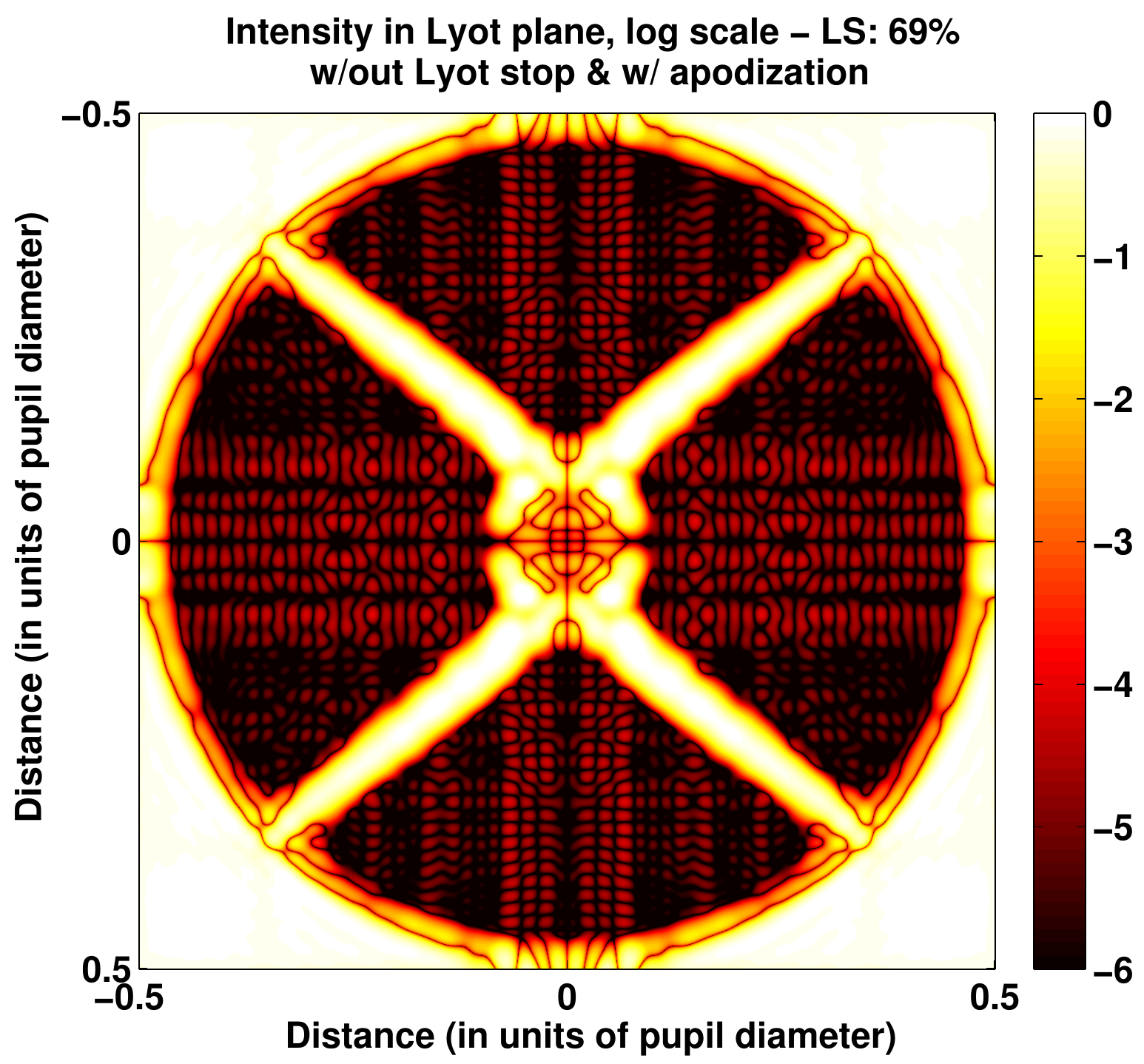}} \subfigure[]{\includegraphics[width=0.5\columnwidth]{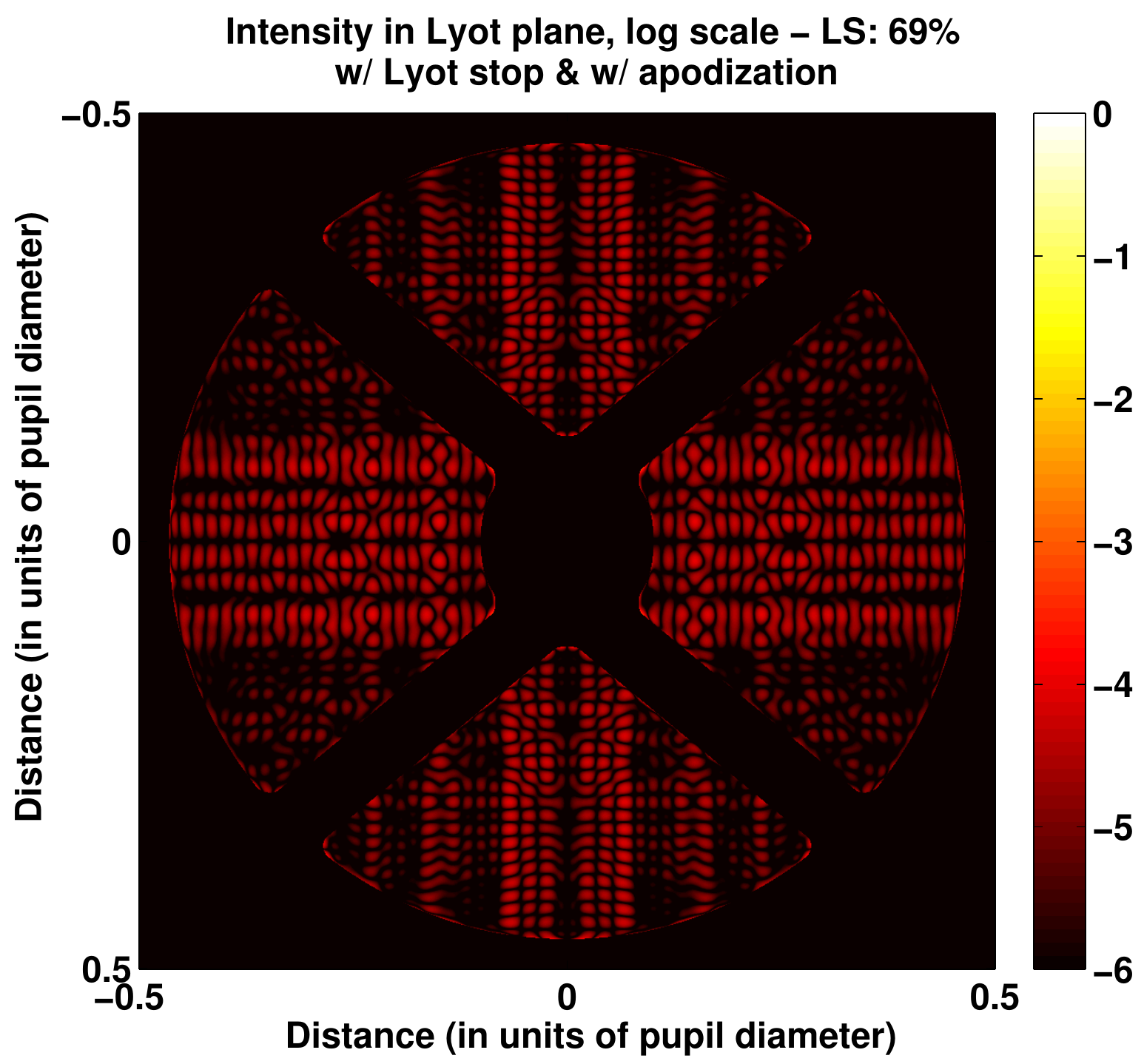}} \\
\end{tabular}
\caption{Log scale intensities in the Lyot plane (a) without the Lyot stop and without apodization, (b) with the Lyot stop and without apodization, (c) without the Lyot stop and with apodization, and (d) with the Lyot stop and apodization. Mask \#4 is used here. It is designed for an 69\% transmission Lyot stop.}
\label{LyotPlane69}
\centering
\end{figure}

The effect of the apodized 4QPM on the spatial distribution of the energy density in the Lyot plane is showed in Fig.\ref{LyotPlane69} for mask \#4. On-axis light is concentrated outside the transmissive area of the Lyot stop and is thus almost entirely blocked by it. On the contrary, in the unapodized case, light is spread in these regions and the extinction of the star is rather poor.

\begin{figure}[]
\centering
\begin{tabular}{cc}
\subfigure[]{\includegraphics[width=0.5\columnwidth]{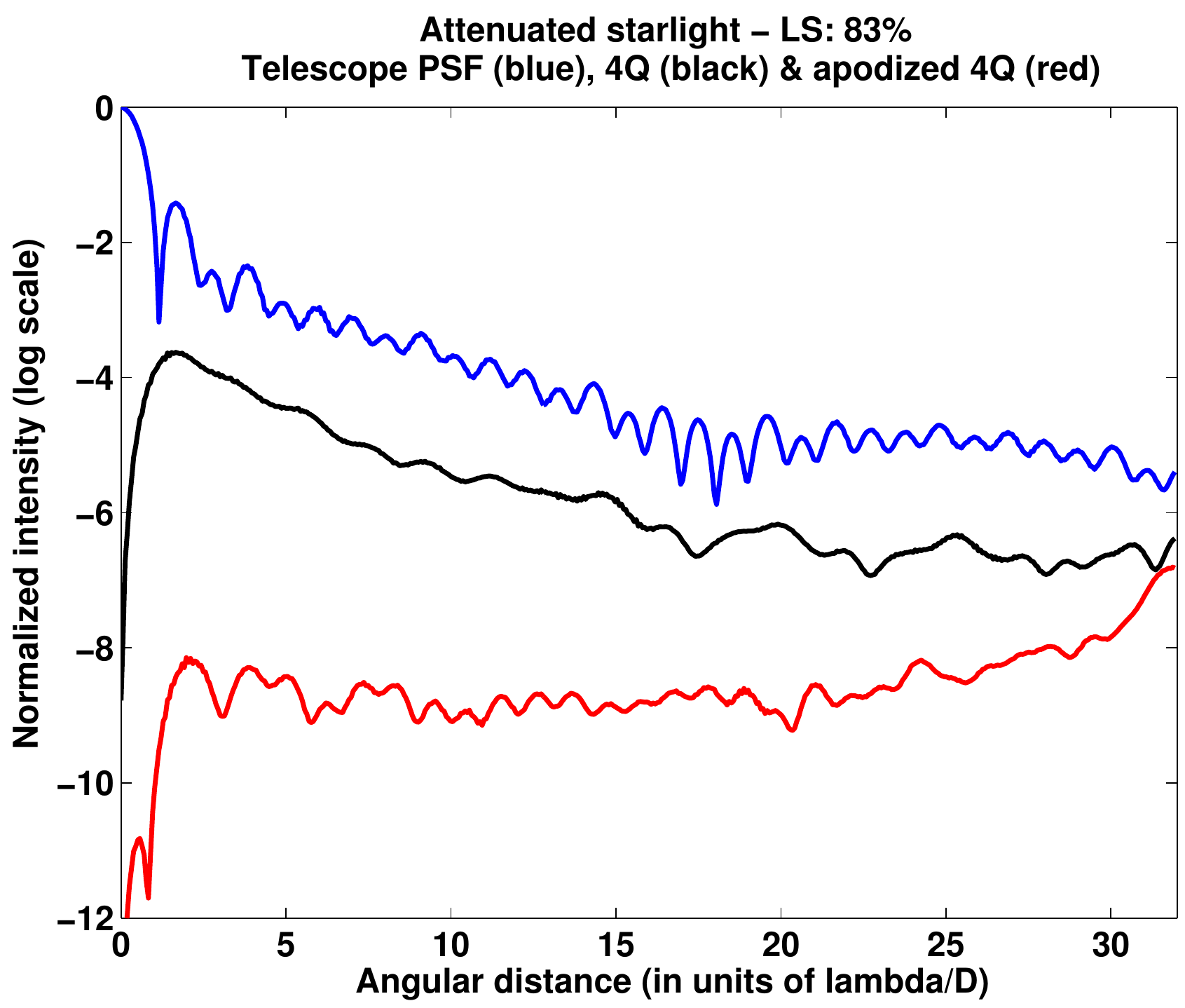}} \subfigure[]{\includegraphics[width=0.5\columnwidth]{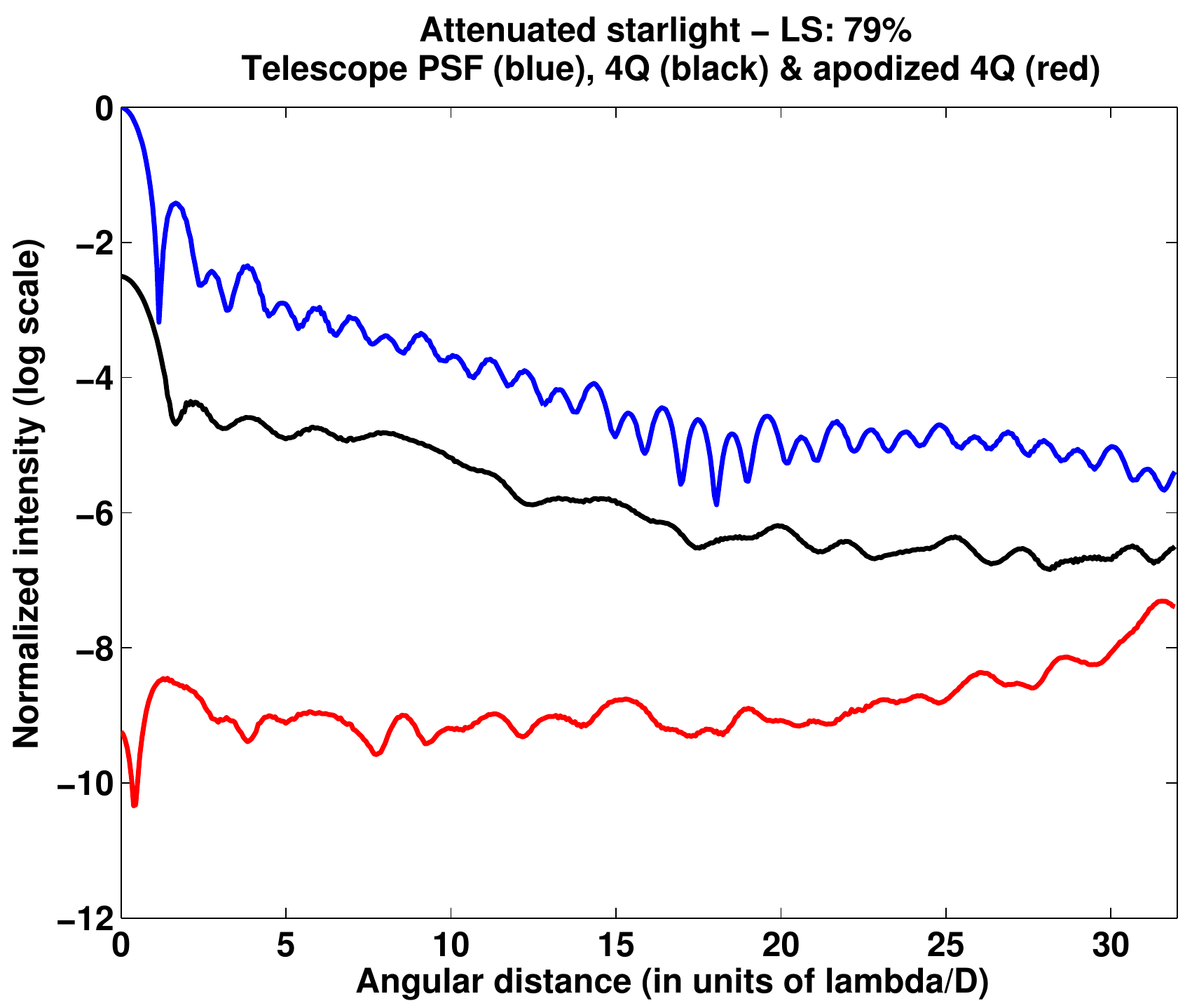}} \\
\subfigure[]{\includegraphics[width=0.5\columnwidth]{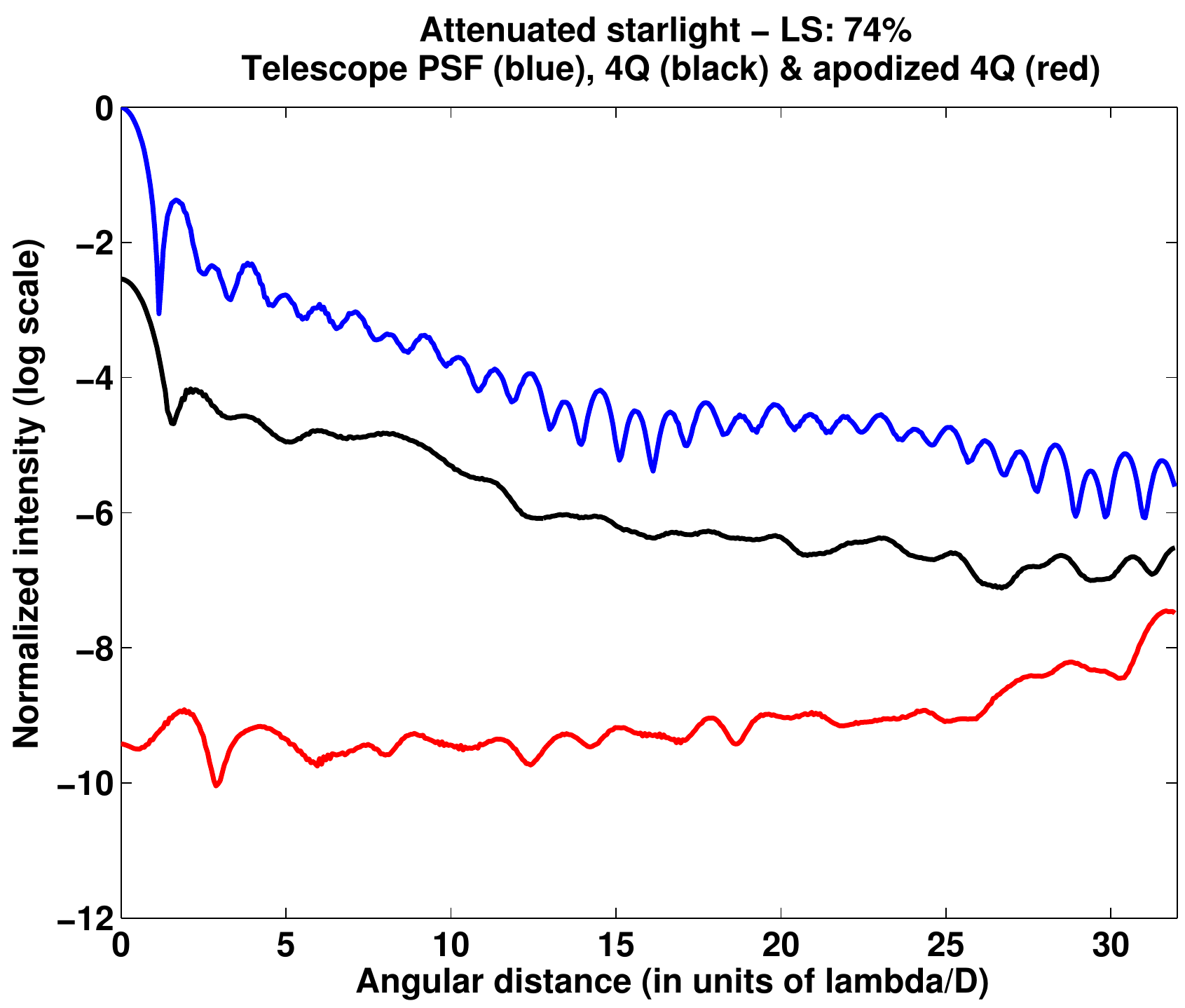}} \subfigure[]{\includegraphics[width=0.5\columnwidth]{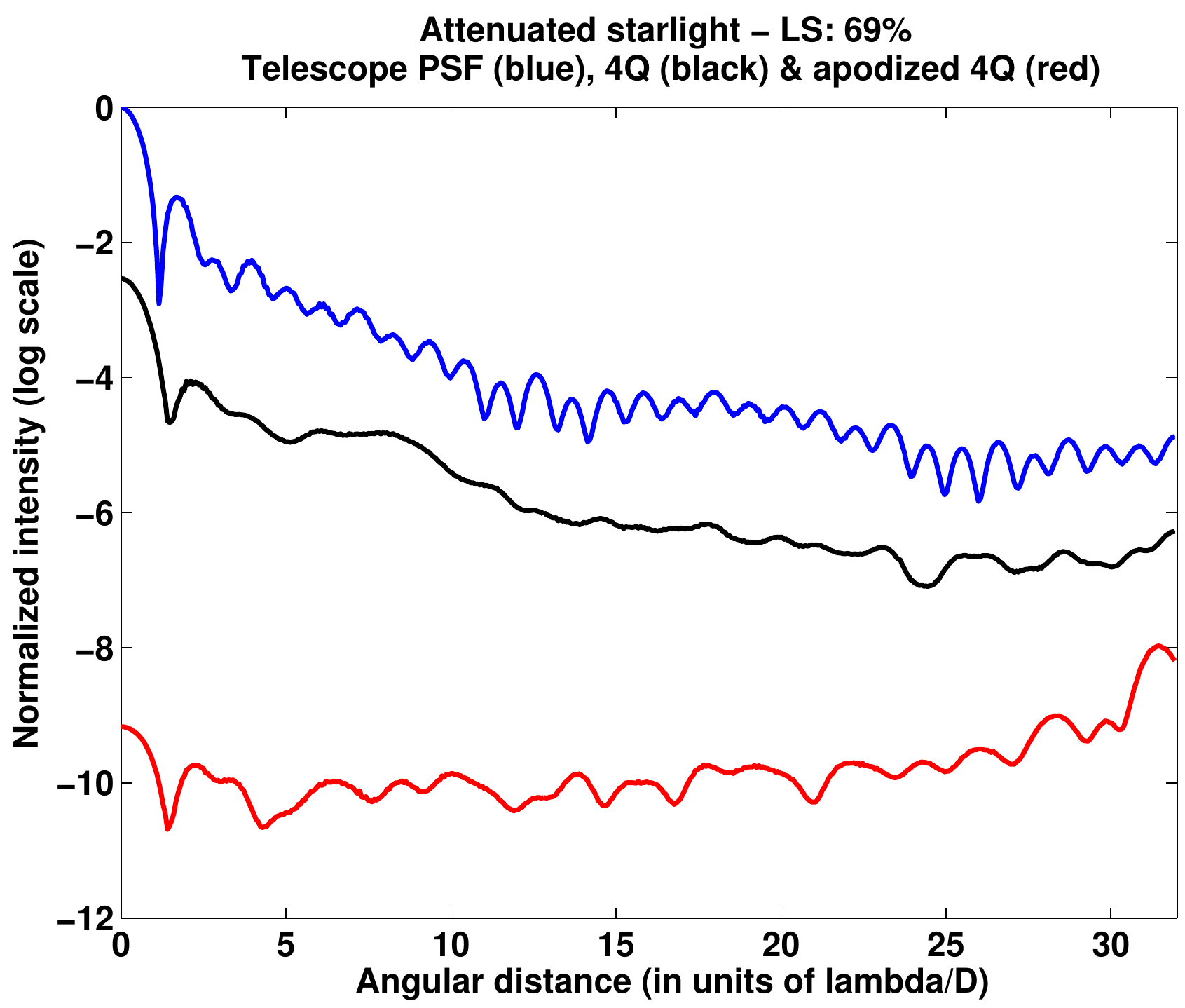}} \\
\subfigure[]{\includegraphics[width=0.5\columnwidth]{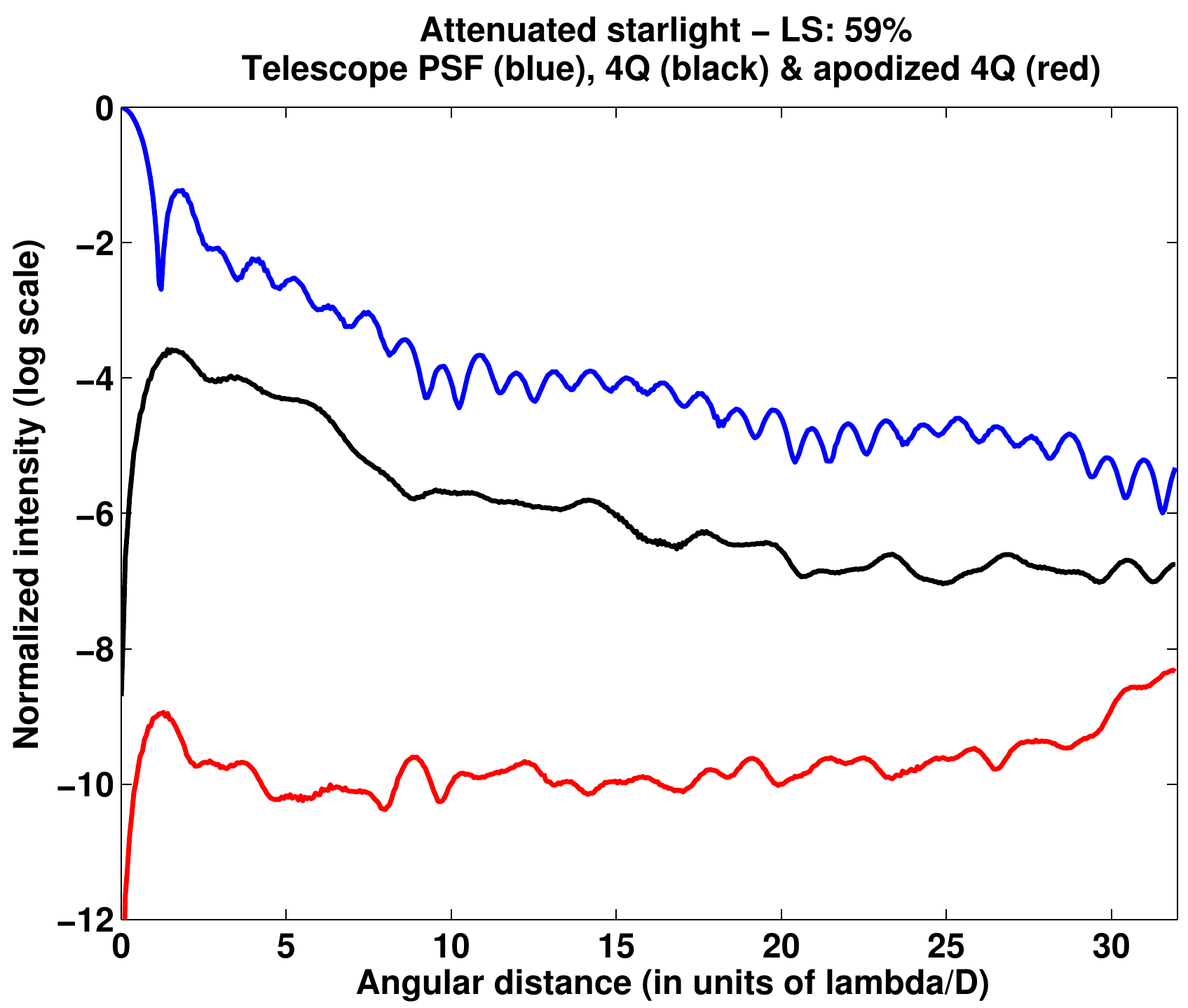}} 
\end{tabular}
\caption{Azimuthally averaged contrast obtained with the apodizers designed a Lyot stop with a transmission of (a) 83\%, (b) 79\%, (c) 74\%, (d) 69\%. In each figure the blue line denotes the PSF of the telescope (including the presence of the Lyot stop), the black line corresponds to the PSF of the 4QPM without apodization, and the red line corresponds to the PSF of the apodized 4QPM.}
\label{Contrast4Q}
\centering
\end{figure}

Radial averaged profiles of the PSFs are displayed in Fig.\ref{Contrast4Q} for the five apodizers (two-dimensional contrast plots were averaged azimuthally so as to be represented in one dimension only). A contrast floor lower than $10^{-8}$ and higher than $10^{-10}$ is obtained with each mask, the lowest ones corresponding to the least transmissive Lyot stops. For comparison, the PSF of the unapodized 4QPM is also showed. A gain of four to five orders of magnitude is provided by using an apodizer.

A comparison of these results with those displayed in Fig.\ref{FiniteMaskSize}, makes it clear that the apodizer does not only help restoring the contrast performance of the 4QPM. While a maximum contrast of $10^{-7}$ should have been obtained for mask size $L$ of 32 $\lambda/D$, maximum contrast as low as $10^{-9}$ are seen in the PSF showed in Fig.\ref{Contrast4Q} (the maximum is looked for over an angular range that extends between 0 and 24 $\lambda/D$, as done it sec.\ref{ImpactSize}). Hence, in addition to limiting the impact of the central obstruction, the apodizer helps mitigate the diffraction effects due to the finite size of the phase mask.

\begin{figure}
\centering
\begin{tabular}{ll}
\subfigure[]{\includegraphics[width=0.45\columnwidth]{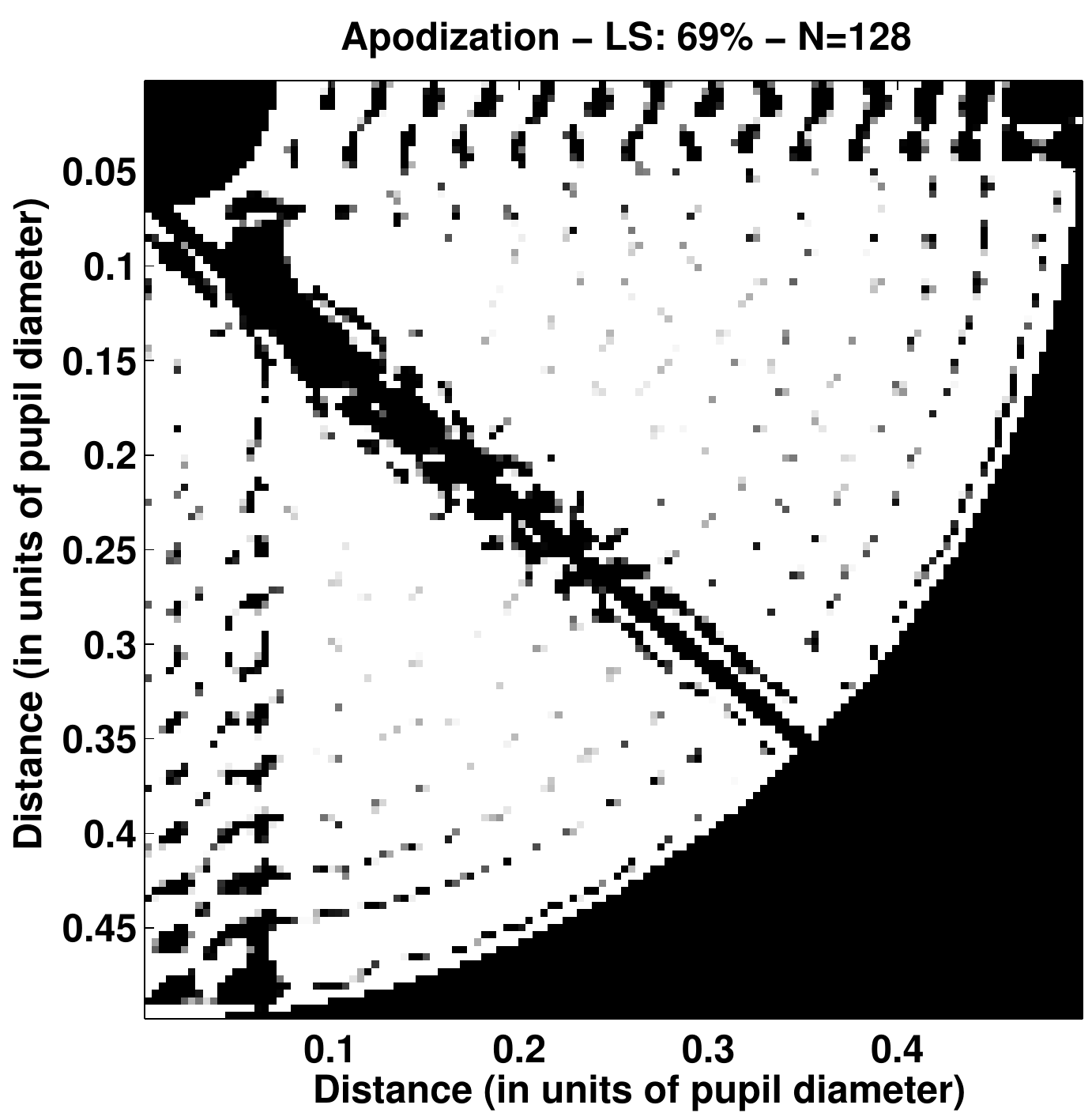}} \qquad \subfigure[]{\includegraphics[width=0.45\columnwidth]{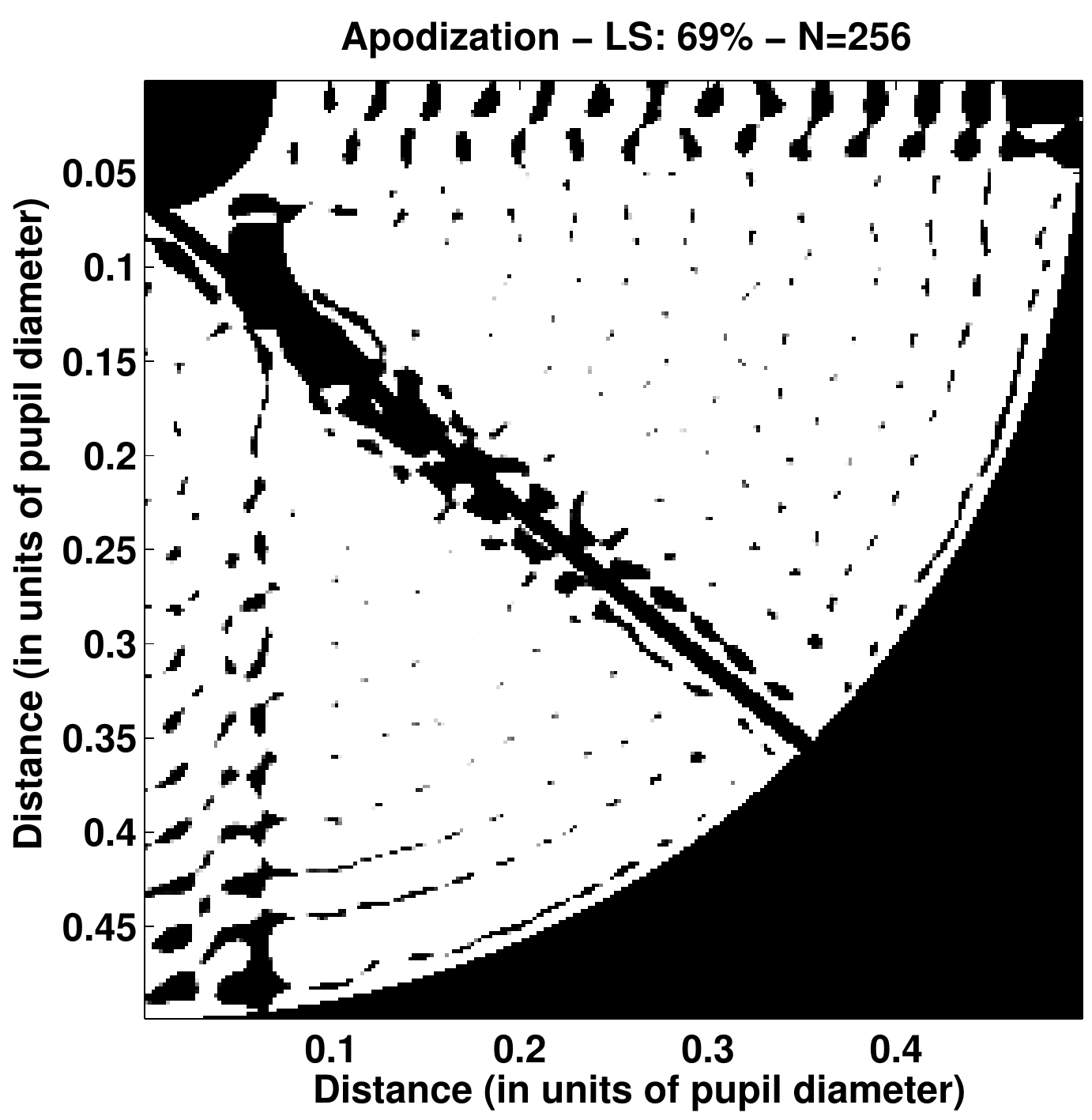}} \\
\end{tabular}
\caption{Transmitted intensities of two apodizers optimized over arrays of different size, for the 4QPM at the VLT and an 69\% transmission Lyot stop. Only one quadrant of the pupil plane is showed. The  number of points in these arrays is (a) 128 by 128 and (b) 256 by 256.}
\label{CompSize}
\end{figure}

It is important to notice that these PSFs were computed using the 'raw' transmission of the apodizers. While the transmissions of the masks tend to be binary, a 5 to 7\% $R_{bin}$ fraction of the pupil area exhibits a transmission greater than 1\% and smaller than 99\% (see Tab.\ref{Table4Q-VLT}). Fig.\ref{CompSize} compares the result of two optimizations that differ by the size of the arrays used to discretize the apodizer in plane A: the first one has 128 by 128 points while the other has 256 by 256 points. The apodizers' transmissions are similar, but there are more than 3 times less points with a non-binary transmission in the larger array ($R_{bin} \approx 1.8\%$). This is not unexpected as it has already been noticed in 2D shaped pupil optimization problems.

Artificially rounding the transmission of the apodizers has a significant impact on the contrast. In the case of mask \#4, as displayed in Fig.\ref{CompSize}, the contrast floor is between $10^{-7}$ and $10^{-6}$. However, the apodizer computed for the same Lyot stop transmission, but over four times as many points, gives a lower contrast that remains between $10^{-8}$ and $10^{-7}$ beyond 3$\lambda/D$, and that does not exceed $10^{-6.6}$ below this angular distance.

As it is summarized in Tab.\ref{Table4Q-VLT}, the respective transmissions $T_{A}$ of the five apodizers range from 92\% to 16\%. These transmissions are defined in the same way as the Lyot stops transmissions were previously defined, as the ratio of the total intensity going through the apodizers with the total intensity going through the initial aperture of the telescope.

The maximum throughput $T_{max}$ of the system ranges from 64\% to 5\%. $T_{max}$ is not computed as the product of $T_{LS}$ and $T_{A}$, as it is the case with most coronagraphs. To do so, one must assume that the focal plane mask does not have a significant impact on the planet light transmission, which fails to be true in the case of the 4QPM. Instead $T_{max}$ is measured by injecting an off-axis source in the instrument and measuring its attenuation in the final image plane.

\begin{figure}
\centering
\begin{tabular}{cc}
\subfigure[]{\includegraphics[width=0.4\columnwidth]{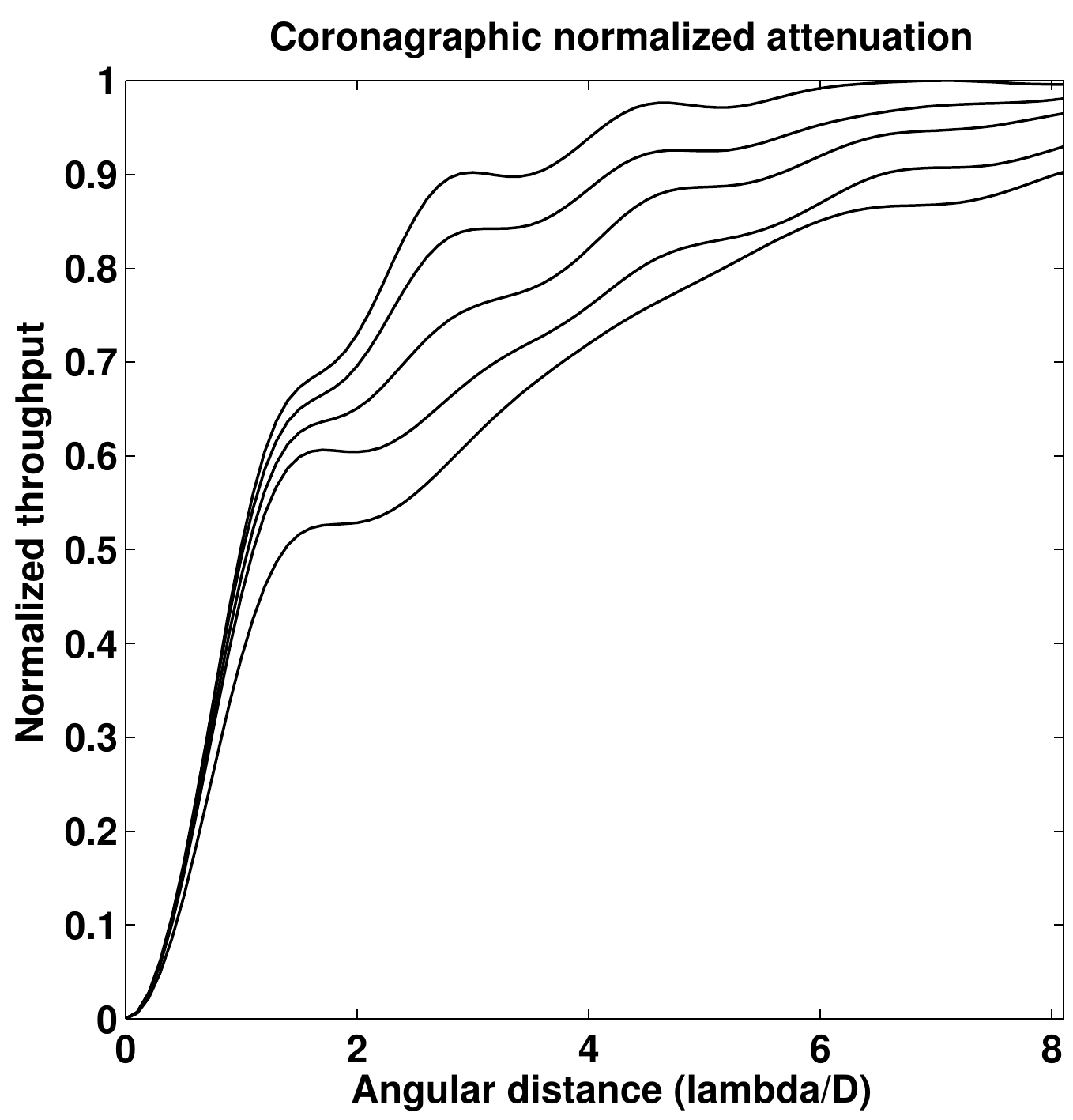}} \qquad \subfigure[]{\includegraphics[width=0.45\columnwidth]{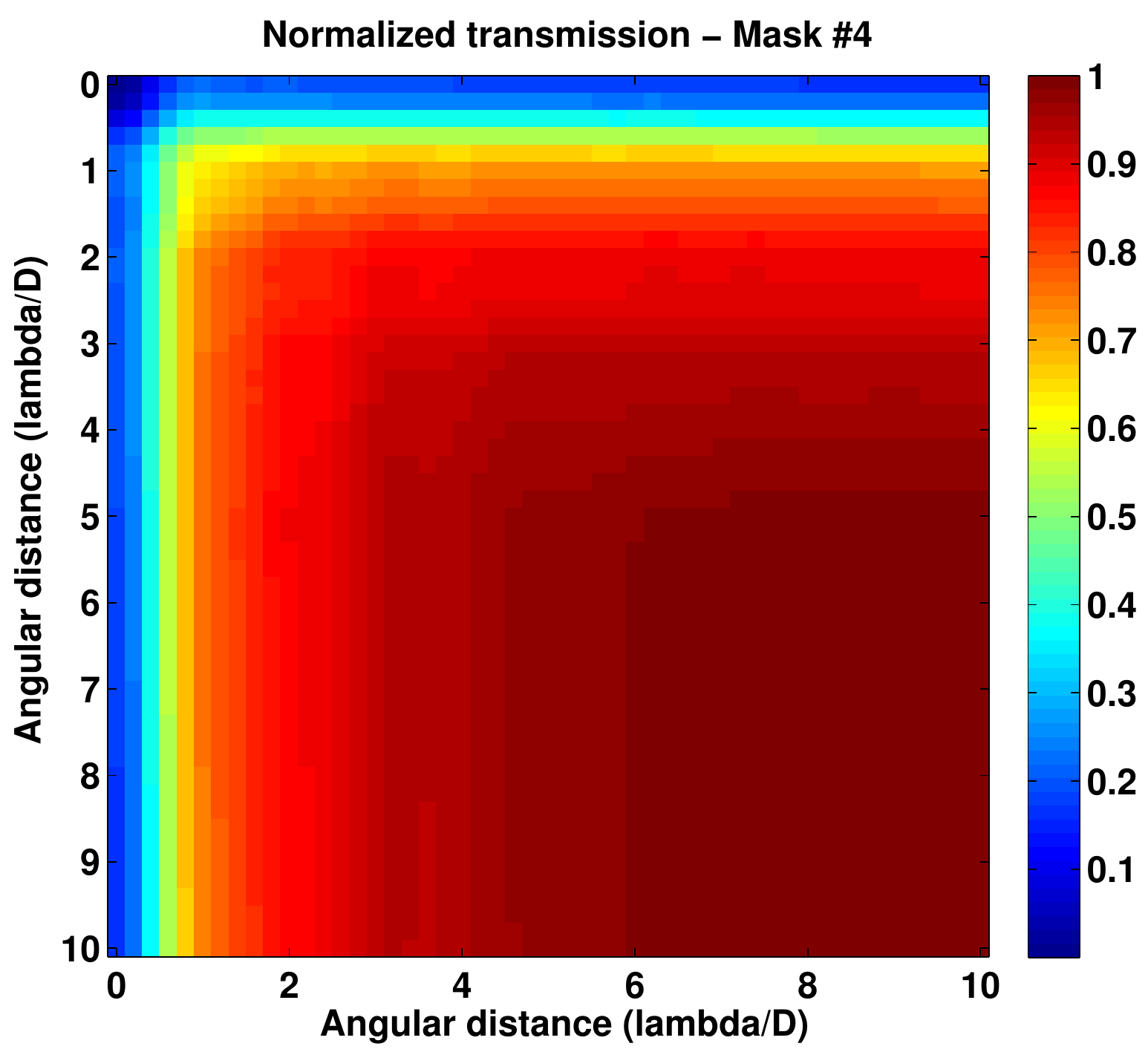}} \\
\end{tabular}
\caption{(a): Normalized transmission of the apodized 4QPM for the five different Lyot stops, and for a planet located, along the diagonal, between 0 and 8 $\lambda/D$. Sorted by increasing transmission, the curves correspond to the 83\%, 79\%, 74\%, 59\% and 69\% transmission Lyot stops. (b): two-dimensional map of the normalized transmission in the case of mask \#4 (Lyot stop transmission: 69\%).}
\label{Transmissions}
\end{figure}

Because of the phase mask, the throughput $T(u,v)$ varies across the image plane. Its maximum, $T_{max}$, is found by moving the planet at equal distance from the 4QPM phase transitions, hence along a diagonal. Fig.\ref{Transmissions} shows how the normalized throughput $T(u,v)/T_{max}$ changes along the diagonals of the masks, and in general in the image plane (but only in the specific case of mask \#4).

\begin{figure}[]
\centerline{\includegraphics[width=0.8\columnwidth]{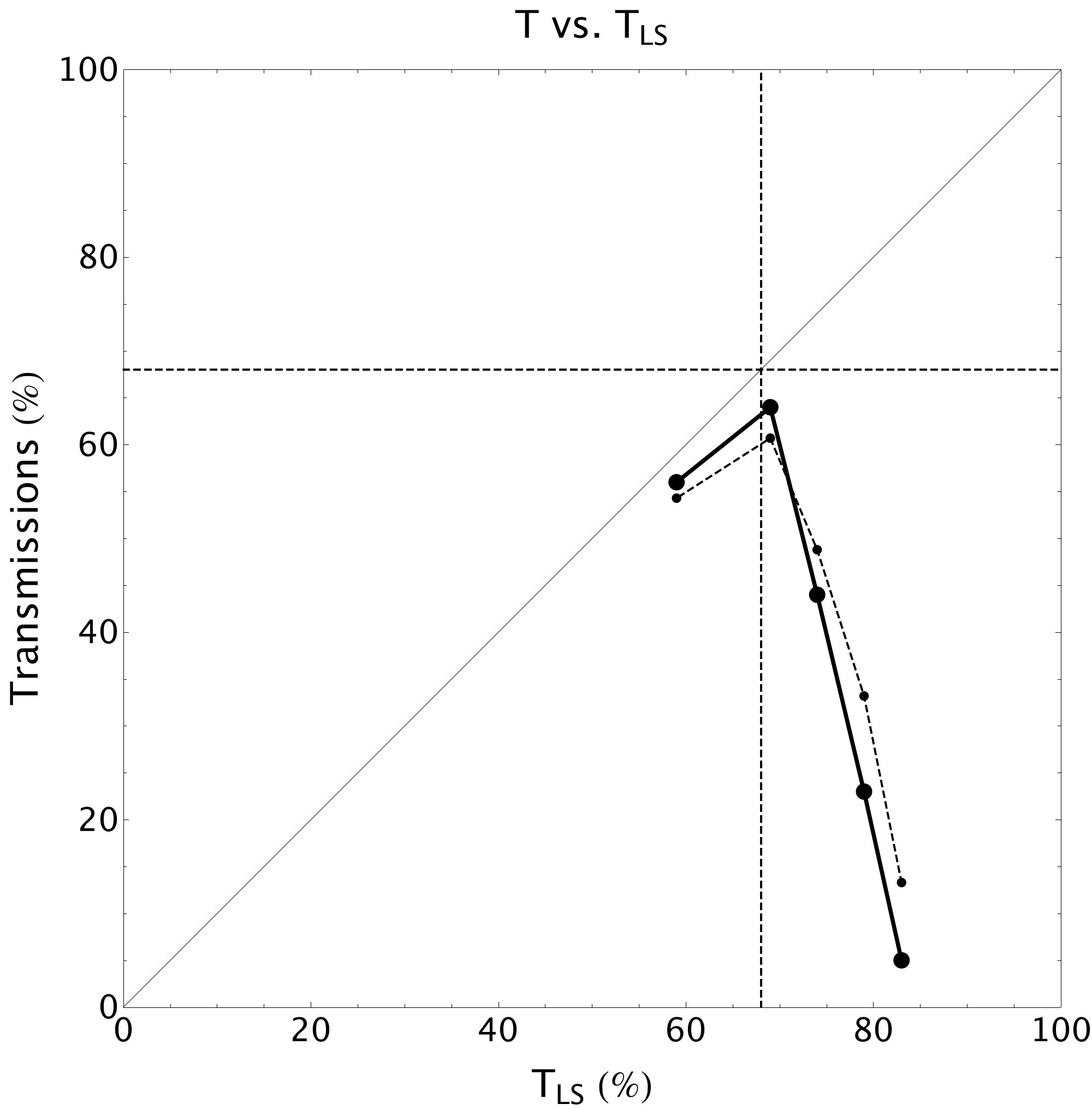}}
\caption{Evolution of the system's throughput $T_{max}$ as a function of the Lyot stop transmission $T_{LS}$. The vertical and horizontal dashed lines indicate the coordinates of the intersection point between the $L=L_{LS}$ line and the linear trend of the four points of the solid curve plotted at the right side of the figure. The dashed curve represents the evolution of the product $T_{LS} \times T_{A}$ as function of $T_{LS}$.}
\label{TransmissionGraph}
\end{figure}

As it is illustrated in Fig.\ref{TransmissionGraph}, to each Lyot stop corresponds a different value of $T_{max}$, which goes through a maximum of 64\% with mask \#4 ($T_{LS}=69\%$). This maximum value can be roughly estimated by extrapolating from a linear fit based on the evolution of $T_{max}$ with $T_{LS}$. The line that can be drawn crosses the line $T_{max}=T_{LS}$ for $T_{LS}\approx 68\%$. Since the system's maximum throughput $T_{max}$ cannot be larger than the Lyot stop's transmission $T_{LS}$, $T=68\%$ can be thought as an upper-bound estimate of the maximum value of $T_{max}$ for this specific telescope architecture, and this family of Lyot stops. Still in Fig.\ref{TransmissionGraph}, the product of $T_{LS}$ and $T_{A}$ is reproduced, as a function of $T_{LS}$. It can be noticed that this product overestimates the value of $T_{max}$ for masks \#1, 2 and 3 and underestimates it for masks \#4 and 5.

The system's maximum throughput $T_{max}$ is however not the only metric one should refer to to evaluate the system's performance. One must also consider the angular resolution of the instrument. %The Lyot stop may degrade it: the 69\% transmission Lyot stop has for instance an effective diameter that is 92\% that of the original diameter, and a 50\% transmission Lyot stop would lead to a 16\% loss in angular resolution (diameters $D_{LS}$ are given for every Lyot stop in Tab.\ref{Table4Q-VLT}). 
Using the information displayed in Fig. \ref{Transmissions} and following \cite{Guyon2006}, the effective IWAs range between 0.99 and 1.37 $\lambda/D$ (see Tab.\ref{Table4Q-VLT}). The smallest IWA corresponds to the 59\% transmission Lyot stop, and the largest IWA to the 83\% transmission Lyot stop.

The time required in the case of the larger array equals 118h, which is 3.3 times longer than the time required for the smaller array. This is in accordance with our estimate of the complexity, as the ratio of the complexities is 3 in this case ($N_{1}=256$, $N_{2}=128$).

\begin{table}
\centering
\begin{tabular}{cccccc}
Mask \# & $T_{LS}$ & $T_{A}$ & $T_{max}$ & $IWA$ & $R_{bin}$ \\
1 & 83\% & 16\% & 5\% & 1.37 $\lambda/D$ & 6\% \\
2 & 79\% & 42\% & 23\% & 1.10 $\lambda/D$ & 7\% \\
3 & 74\% & 66\% & 44\% & 1.06 $\lambda/D$ & 7\% \\
4 & 69\% & 88\% & 64\% & 1.01 $\lambda/D$ & 6\% \\
5 & 59\% & 92\% & 56\% & 0.99 $\lambda/D$ & 5\%
\end{tabular}
\caption{Main parameters of the five different apodizers optimized for a 4QPM coronagraph. Each mask corresponds to a specific Lyot stop, with a transmission $T_{LS}$. $T_{A}$ is the transmission of the apodizer. This results in the system's total throughput $T$. The IWA of the coronagraph is given in units of $\lambda/D$. Finally, $R_{bin}$ is the ratio between the non-binary surface of the mask and the total surface of mask. The non-binary elements are defined as having a transmission greater than 1\% and less than 99\%.}
\label{Table4Q-VLT}
\end{table}

%\begin{figure}[]
%\centering
%\begin{tabular}{cc}
%\subfigure[]{\includegraphics[width=0.5\columnwidth]{VLT-4Q-LS83-LS0-AP0.pdf}} \subfigure[]{\includegraphics[width=0.5\columnwidth]{VLT-4Q-LS83-LS1-AP0.pdf}} \\
%\subfigure[]{\includegraphics[width=0.5\columnwidth]{VLT-4Q-LS83-LS0-AP1.pdf}} \subfigure[]{\includegraphics[width=0.5\columnwidth]{VLT-4Q-LS83-LS1-AP1.pdf}} \\
%\end{tabular}
%\caption{Log scale intensities in the Lyot plane (a) without the Lyot stop and without apodization, (b) with the Lyot stop and without apodization, (c) without the Lyot stop and with apodization, and (d) with the Lyot stop and apodization. Mask \#1 is used here. It is designed for an 83\% transmission Lyot stop.}
%\label{LyotPlane83}
%\centering
%\end{figure}

\begin{figure}[]
\centerline{\includegraphics[width=0.8\columnwidth]{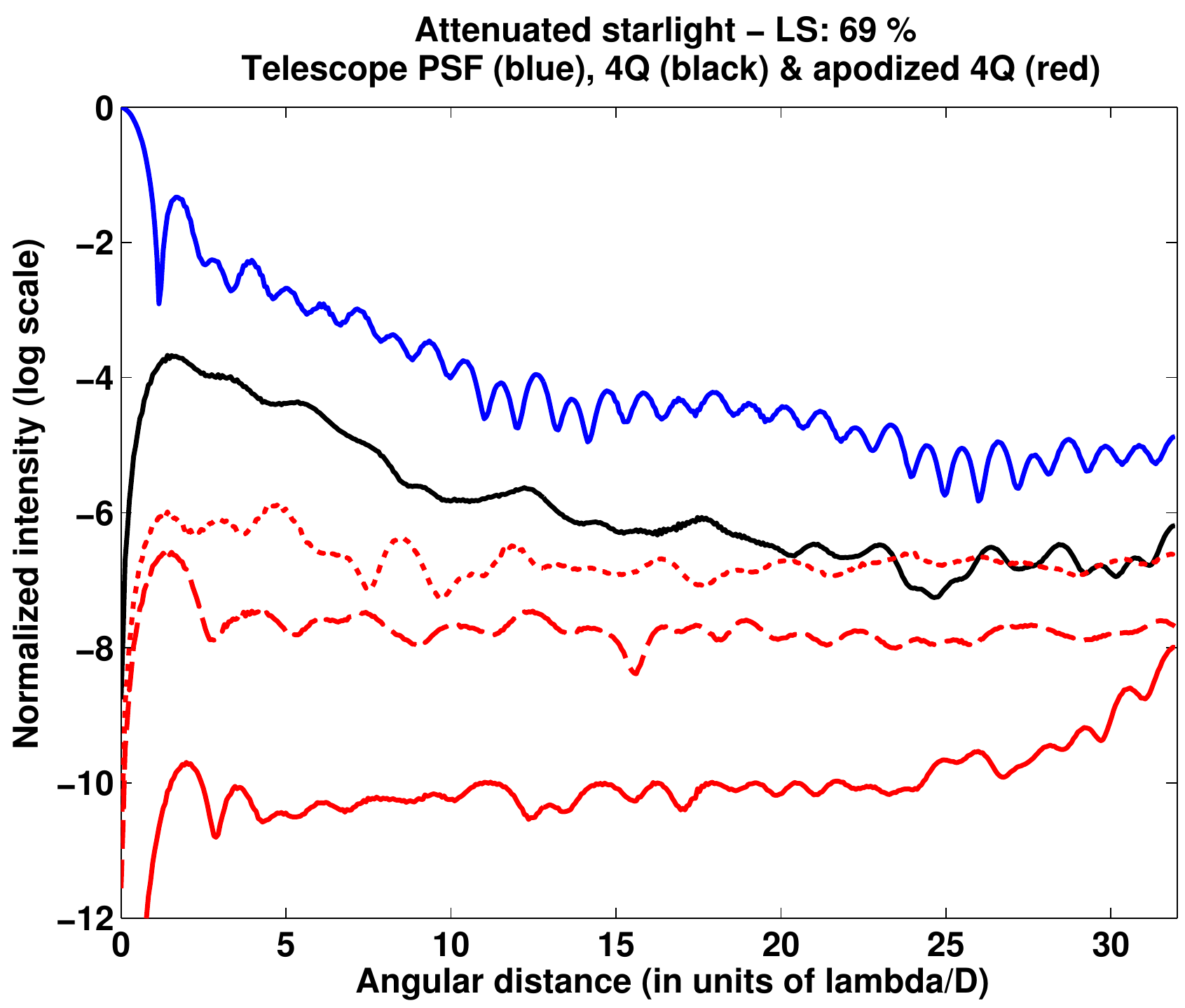}}
\caption{Azimuthally averaged contrast plots obtained with the apodizer designed for a 69\% transmission Lyot stop. As in Fig.\ref{Contrast4Q}, the blue line denotes the PSF of the telescope (including the presence of the Lyot stop) while the black line corresponds to the PSF of the 4QPM without apodization. The three red lines correspond to the PSF of the apodized 4QPM. The solid line is used for the PSF of the 'raw' mask (i.e. non-binary), and the dashed and dotted lines are used respectively for the PSFs of the binary masks computed over 256 by 256 points and 128 by 128 points.}
\label{PSFLargeSmall}
\end{figure}

\section{Discussion \& Conclusion}
\label{Conclusion}

Both the original four-quadrant phase mask coronagraph and the vector vortex coronagraph see their performance reduced with centrally obstructed apertures such as the VLT's, and the contrast they provide is only 1 to 2 orders of magnitude below the PSF of the telescope. On the contrary, when used with clear circular apertures, these coronagraphs could provide contrast as low as $10^{-10}$ at only 1$\lambda/D$ from the star.

Solutions have been proposed in the past that consist in either using a circular subaperture, or using a two-stages optical layout. The first solution leads to an important resolution loss and an even higher transmission loss. The second solution offers a more satisfying result, but requires twice as many optical components, which can limit the system's performance because of alignment issues, additional wavefront aberrations and chromaticity effects.

It was shown that optimal apodizers can help restore the performance of these coronagraphs when used with an on-axis telescope, and that the diffraction effects caused by the finite size of the focal phase mask are also mitigated thanks to the apodizer. In addition to the already existing phase mask and Lyot stop, an apodizer is located in an upstream pupil plane. No other additional optical components are required. Any aperture can be used as a basis for an optimization, although symmetric apertures have the advantage of reducing the number of points and thus the complexity of the computations.

The numerical optimization of the apodizer's transmission can be done in two dimensions as it is already done for shaped pupils. The complexity of the problem is however different whether a 4QPM or a VPM is used. Different cases require different computations methods. In the case of the 4QPM, the direct computation of the electric field in the Lyot plane appears to be a better choice. Indeed, sampling errors in plane B are avoided since the electric field need not be computed explicitly in this plane. In the case of the VPM however, the computation of the same electric field should better be done through two successive Fourier transforms (from pupil plane A to image plane B, and then to pupil plane C). While 1.5 days are necessary to compute an apodizer over a 128 by 128 matrix for the 4QPM, the same optimization would take about 3 to 9 times longer for the VPM, depending on the sampling resolution of plane B.

The apodizers that are presented here are designed only for the 4QPM. A VLT-like aperture was chosen and different Lyot stop sizes were considered. For each Lyot-stop, an apodizer is found by solving a linear convex optimization problem. The total throughput of the system is a function of the Lyot stop's transmission. It goes through a maximum for which an upper-bound of about 68\% is found for a 68\% transmission Lyot stop (and in the particular case of the VLT). In practice, i.e. when optimizing an apodizer for this specific Lyot stop, the system's throughput that is found is 64\% and is given by an apodizer with an 89\% transmission, associated to a 69\% transmission Lyot stop.

As observed previously with fully optimized shaped pupils, the masks that have been computed tend to have a binary transmission: only 5 to 7\% of the pupil surface has a transmission higher than 1\% and smaller than 99\%. This number goes down to 1.7\% when the size of the array is doubled, from 128 by 128 to 256 by 256. It is very likely that masks optimized over even larger arrays would have an even smaller non-binary transmission ratio. Computing a mask over a 512 by 512 array would take an estimated time of 15 days. In the case of full two-dimensional shaped pupil optimization, array's length of half a thousand points have proven to be large enough to obtain quasi-binary apodizers whose transmission, when artificially rounded, provides contrast down to $10^{-9}$\citep{Carlotti2011}. Contrast values are significantly increased when artificially rounding the transmission of the apodizers: up to about $10^{-6}$ for a 128 by 128 mask, and up to about $10^{-7}$ for a 256 by 256 mask instead of a few $10^{-10}$ for mask \#4. Optimizing masks over larger arrays is mandatory, and this point will be addressed in the near future, thanks to a new computer, dedicated to weeks-long computations.

Binary masks can efficiently be manufactured using photolithography: a thin metallic coating can be deposited on a glass substrate and then selectively etched. The substrate adds chromaticity and wavefront aberrations. If the mask is structurally connected, it can be etched out of a thin silicon wafer coated with metal, eliminating these disadvantages.

Gray apodizers have also been produced, either by depositing on a substrate a metallic layer with a spatially varying thickness, or by using high energy beam sensitive (HEBS) glass. These apodizers are however considered too chromatic \citep{Sivaramakrishnan2009}.

Apart from mask \#1, every mask computed for the VLT-like aperture would require the use of a substrate. While a free-standing apodizer would certainly have its advantages, the apodized coronagraphs of both SPHERE and GPI use substrate-based apodizers. As presented in \cite{Enya2008}, substrate-based shaped-pupils with anti-reflection coatings on both sides have achieved $8 \times 10^{-8}$ contrast in the visible. SPHERE and GPI both work in the near-infrared where, given the same surface aberrations, a substrate would create weaker wavefront aberrations than in the visible, which could potentially lead to even lower contrast values.

The possibility also exists that MEMS or MOEMS devices could be used as adaptive binary apodizers, either in transmission (using a technology close to the one used in JWST's micro-shutter arrays) or in reflection (such as the micro-mirror arrays used in many video-projectors). Apart from offering substrate-free apodizers, it would also allow a direct amplitude control of the wavefront.

Apodizers optimized for a 4QPM could potentially complement SPHERE's 4QPM coronagraphs, increasing the achievable contrast while offering the same very small resolution.

%Computing apodizers for JWST-MIRI appears to be possible, although the asymmetry of JWST's aperture would lead to higher complexities and thus longer computation times.

No apodizer could yet be optimized for the vector vortex coronagraph. The VVC has however a substantial advantage over the 4QPM since it does not reduce the discovery area.
Both the charge 2 VVC and the 4QPM are very sensitive to finite stellar size and jitter. This sensitivity can be reduced if a charge 4 VVC or an 8 octants mask is used instead, although the 8 octants reduces the discovery area even more than the 4QPM does.

There is no inherent reason that makes the optimization of apodizers for the VVC impossible. Shortening the computation time would however greatly help this process. To that end it will be necessary to either simplify the problem further, possibly by assuming circular-symmetric apertures and apodizers, or to solve the optimization problem in a more efficient manner, for example by taking advantage of the multiple cores of a computer.

Finally, the joint optimization of the apodizer and the Lyot stop is also a possibility that will be explored. Though more computationally intensive, it would de facto maximize the transmission of the system.

%\Large{acknowledgements}
%The author is thankful to the anonymous referee for his/her meaningful suggestions, and also to Dr. Laurent Pueyo and Dr. Dimitri Mawet for their encouragements, and for the fruitful discussions that he shared with them as well as with Mary-Ann Peters.

\bibliographystyle{aa} % style aa.bst
\bibliography{BIB} % your references Yourfile.bib

\end{document}